\documentclass[12pt]{article}
\usepackage{a4wide,epsfig,amsmath,amssymb,cite}
\usepackage{scalefnt}
\newcommand{\qsla}{q\!\!\!/\,\,}

\newcommand{\logm}[1]{L_{#1}}
\newcommand{\logmbar}[1]{\overline{L}_{#1}}
\newcommand{\logtwo}{\ln2}
\newcommand{\logthree}{\ln3}
\newcommand{\logthreetwo}{\ln^2 3}
\newcommand{\sqrtthree}{\sqrt{3}}
\newcommand{\yh}{y_H}
\newcommand{\yw}{y_W}
\newcommand{\yz}{y_Z}
\newcommand{\yhone}{y_{H,1}}
\newcommand{\yone}{y_1}
\newcommand{\yhbar}{{\overline y}_H}
\newcommand{\ywbar}{{\overline y}_W}
\newcommand{\yzbar}{{\overline y}_Z}
\newcommand{\yhbarone}{{\overline{y}}_{H,1}}
\newcommand{\nyeptwo}{Y^\epsilon}
\newcommand{\lsthree}{\mbox{Ls}_3\left (\frac{2\pi}{3}\right)}

\begin{document}

%%%%%%%%%%%%%%%%%%%%%%%%%%%%%%%%%%%%%%%%%%%%%%%%%%%%%%%%%%%%

\title{\vskip-3cm{\baselineskip14pt
\begin{flushleft}
\normalsize SFB/CPP-05-78 \\
\normalsize TTP05-26 \\
%%%\normalsize DESY 05-xxx
\end{flushleft}}
\vskip1.5cm
Two-Loop ${\cal O}(\alpha\alpha_s)$ Corrections to the
On-Shell Fermion Propagator in the Standard Model}
\author{\small D. Eiras, M. Steinhauser\\
  {\small\it Institut f{\"u}r Theoretische Teilchenphysik,
    Universit{\"a}t Karlsruhe}\\
  {\small\it 76128 Karlsruhe, Germany}\\
}

\date{}

\maketitle

\thispagestyle{empty}

\begin{abstract}
In this paper 
we consider mixed two-loop electroweak corrections to the top quark
propagator in the Standard Model.
In particular, we compute the on-shell renormalization constant
for the mass and wave function, which constitute building blocks for
many physical processes.
The results are expressed in terms of master integrals. For the latter
practical approximations are derived. 
In the case of the mass renormalization constant we find agreement with the
results in the literature.

\medskip

\end{abstract}

\newpage

%- }}}
%- {{{ Introduction:

\section{Introduction}

\indent

The outstanding precision reached at the CERN LEP and SLAC SLC triggered many
higher order calculations in the Standard Model (SM) of 
particle physics. In particular, it happened for the first time that
the experimental results were sensitive to the weak part of the SM.
Since at LEP and SLC real top quarks could not be produced,
the main emphasis of the theoretical investigations 
was put on processes with 
light quarks as external particles. Heavy particles like the
top quark only appeared virtually in intermediate states.
In many applications the masses of the light quarks can be neglected as
compared to the other mass scales, which results in a significant
simplification of the resulting mathematical expressions.

In a future International Linear Collider (ILC)~\cite{ilc} 
the center-of-mass energy is high enough to produce top quarks.
The expected experimental precision requires on the theoretical
side the inclusion of higher order corrections --- both for the QCD and
the electroweak sector of the SM. 
This is particularly true for the threshold production of top quark pairs, 
where the theoretical uncertainties of the 
second order QCD corrections are still
significant~\cite{Hoang:2000yr}. Thus, next to 
third-order QCD calculations
(see, e.g., Refs.~\cite{KniPenSmiSte,PenSte,KPSS2,Hoa}), also
electroweak corrections have to be investigated.

\begin{figure}[b]
  \begin{center}
    \begin{tabular}{c}
    \epsfig{figure=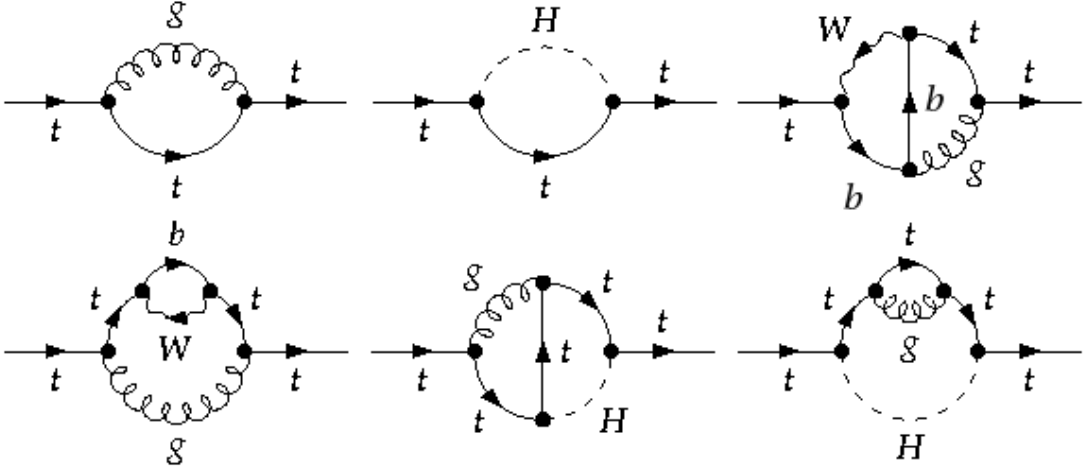,width=36em}
    \end{tabular}
    \parbox{14.cm}{
      \caption[]{\label{fig::sample}\sloppy Sample diagrams
        contributing to the top quark propagator up to order $\alpha\alpha_s$.
        Next to the Higgs boson also the gauge bosons $W$ and $Z$ and
        the corresponding Goldstone bosons can be exchanged.
        }}
  \end{center}
\end{figure}

In this paper we focus on the two-loop mixed electroweak/QCD
corrections to the on-shell top quark propagator. Some sample diagrams
are shown in Fig.~\ref{fig::sample}.
The top quark mass renormalization constant has been
evaluated in Ref.~\cite{Jegerlehner:2003py}. The contribution from the 
scalar bosons has been considered
in~\cite{Faisst:2003px,Faisst:2004gn}.
In this paper we confirm the results of 
Ref.~\cite{Jegerlehner:2003py}.
Furthermore, we compute the wave function 
renormalization constant to order $\alpha\alpha_s$.
Practical approximations are derived for the diagrams with internal
$W$ and $Z$ bosons\footnote{Here and in the following we
  consider the $W$ and $Z$ boson always in combination with the
  corresponding Goldstone bosons.}
and both the exact expression and handy
approximations are computed for the Higgs boson mass dependence.

The inverse fermion propagator can be decomposed as
\begin{eqnarray}
  S^{-1}(q) = 
  Z_2^L\left(\qsla - Z_m m + \qsla \Sigma_L(q) + m\Sigma_S\right) L
  + Z_2^R\left(\qsla - Z_m m + \qsla \Sigma_R(q) + m\Sigma_S\right) R
  \,,
  \label{eq::fprop}
\end{eqnarray}
with $R = (1+\gamma_5)/2$ and $L = (1-\gamma_5)/2$.
$m$ is the quark mass and 
$Z_m$ and $Z_2^{L,R}$ are the mass and (left/right) wave function 
renormalization constants, respectively.  
The functions $\Sigma_L, \Sigma_R$ and $\Sigma_S$ in
Eq.~(\ref{eq::fprop}) result from a convenient
decomposition of the self energy $\Sigma(q)$ given by
\begin{eqnarray}
  \Sigma(q) &=&  
  \qsla (R \, \Sigma_R(q^2) + L \, \Sigma_L(q^2) ) + m \, \Sigma_S(q^2)
  \,.
  \label{eq::sigma}
\end{eqnarray}

In the on-shell scheme one requires that
$S^{-1}(q)$ vanishes for $q^2=m^2$ which leads to the following
condition for the on-shell mass renormalization constant
\begin{eqnarray}
  Z_m^{\rm OS} &=&
  1+\left (\Sigma_S(m^2)
  +\frac{1}{2}\left(\Sigma_L(m^2)+\Sigma_R(m^2) \right)\right)
  \,.
\end{eqnarray}
Requiring furthermore that the residuum is 
$-1$, provides a condition
for $Z_2^{L,{\rm OS}}$ and $Z_2^{R,{\rm OS}}$
\begin{eqnarray}
  Z_2^{L,{\rm OS}} &=& -\Sigma_L(m^2) -2 m^2 \left[\Sigma_S^{\prime}(m^2)
  +\frac{1}{2}\left(\Sigma_L^{\prime}(m^2)
  +\Sigma_R^{\prime}(m^2) \right)\right]\, ,
  \nonumber\\
  Z_2^{R,{\rm OS}} &=& -\Sigma_R(m^2) -2 m^2 \left[\Sigma_S^{\prime}(m^2)
  +\frac{1}{2}\left(\Sigma_L^{\prime}(m^2)
  +\Sigma_R^{\prime}(m^2) \right)\right]
  \,.
\label{eq::z2lr}
\end{eqnarray}
In these formulae it is understood that only the real part of the
self energy functions is taken.

In the practical calculation it is convenient to
apply projectors in order to arrive at scalar momentum
integrals. This is achieved via
\begin{eqnarray}
  Z_m^{\rm OS} &=& \mbox{Tr}\left(P_m \Sigma(q)\right)\bigg|_{q^2=m^2}
  \,,\nonumber\\
  Z_2^{L/R,\rm OS} &=& \mbox{Tr}\left(P_2^{L/R}\Sigma(q)\right)\bigg|_{q^2=m^2}
  \,,
\end{eqnarray}
with
\begin{eqnarray}
  {P_m} &=& \frac{\qsla}{4 q^2}+\frac{{\bf 1}}{4 m} \,,
  \nonumber \\  
  {P_2^L} &=& -\frac{\qsla}{2 q^2}{L}
  -2 m^2 \frac{\partial}{\partial q^2} \left ( \frac{\qsla}{4 q^2}
  +\frac{{\bf 1}}{4 m} \right ) \,,
  \nonumber \\  
  {P_2^R} &=& -\frac{\qsla}{2 q^2}{R}
  -2 m^2 \frac{\partial}{\partial q^2} \left ( \frac{\qsla}{4 q^2}
  +\frac{{\bf 1}}{4 m} \right ) \,.
\end{eqnarray}
In these formulae it is understood that the lower-order results are
expressed in terms of the bare parameters.
In what follows we compute $Z_m^{\rm OS}$ and $Z_2^{L/R,\rm OS}$ 
for the top quark neglecting the masses of the light quarks.
Furthermore, we take the Cabibbo-Kobayashi-Maskawa matrix to be
diagonal.

The outline of the paper is as follows:
In the next Section we consider the classes of Feynman diagrams which are
relevant for our calculation and discuss the reduction to a basic set
of master integrals. 
The results for the master integrals appearing in
our calculation are discussed in Appendix~\ref{app::ae_of_masters}.
In Section~\ref{sec::Zm} we present the renormalization constant for
the top quark on-shell mass. In particular we consider the
relation between the $\overline{\rm MS}$ and pole mass and compare our
results with the literature. 
In Section~\ref{sec::Z2} we move on to the 
wave function renormalization constants.
Finally, our findings are summarized in Section~\ref{sec::concl},
which also contains the conclusions.

%- }}}
%- {{{ Reduction to master integrals:

\section{Reduction to master integrals}

\indent

At order $\alpha\alpha_s$ one has to consider about 35
Feynman diagrams contributing to the fermion propagator
(cf. Fig.~\ref{fig::sample} for some sample diagrams).
After the application of the projectors 
the external momentum is set on the mass shell of the
heavy quark, which leads to integrals containing two scales, the quark
mass and the boson mass.
In this Section they are denoted by $m$ and $M$, respectively. 

We generate all one-particle irreducible Feynman diagrams 
contributing to the fermion propagator with {\tt QGRAF}~\cite{Nogueira:1991ex}.
The application of {\tt q2e} and 
{\tt exp}~\cite{Harlander:1997zb,Seidensticker:1999bb}
identifies the topology of the individual diagrams, adopts the
notation and transforms the expressions into {\tt
  FORM}~\cite{Vermaseren:2000nd} notation.

\begin{figure}[t]
  \begin{center}
    \begin{tabular}{c}
      \epsfxsize=35em
      \epsffile[60 350 550 450]{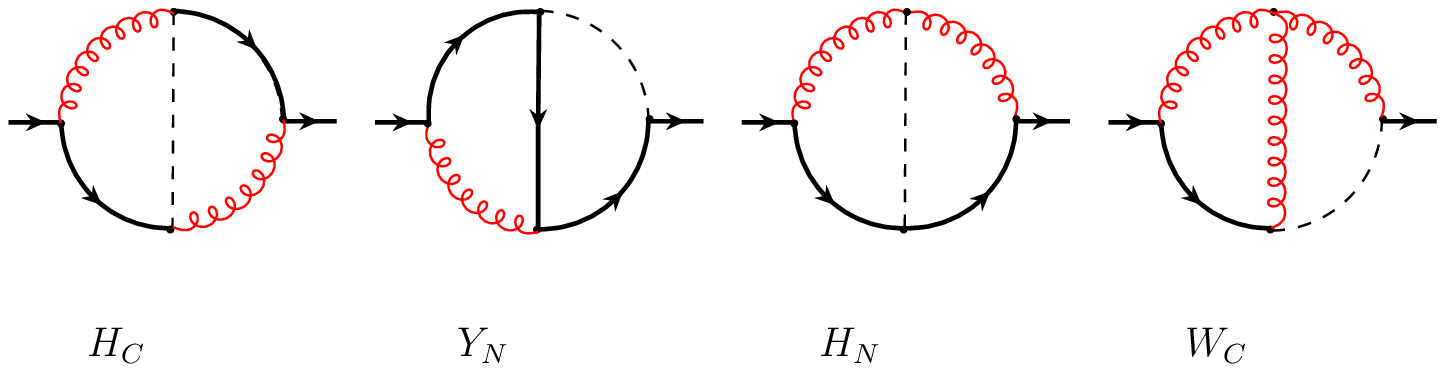}
    \end{tabular}
    \parbox{14.cm}{
      \caption[]{\label{fig::FDclasses}\sloppy 
        Graphical representation of the integral classes defined in
        Eq.~(\ref{eq::FDclasses}). Solid and dashed lines carry
        mass $m$ and $M$, respectively. Curly lines are massless.
        }}
  \end{center}
\end{figure}

It is convenient to map the {\tt QGRAF}
output for each diagram 
with the help of {\tt q2e} and {\tt exp} to one of the
following four classes of 
integrals\footnote{For the application at hand it is possible to
  work with a smaller set of integral types. E.g.,
  in the case of $H^+_C$ we have $n_5\le0$ for the self energies
  considered in this paper. However, in view of
  a future application to on-shell vertices
  it is advantageous to consider a more general set-up.}
\begin{eqnarray}
  \lefteqn{H^+_N(n_1,n_2,n_3,n_4,n_5) =}\nonumber\\&&
  \frac{e^{2\epsilon\gamma_E}}{(i\pi^{d/2})^2} 
  \int\frac{ {\rm d}^d k {\rm d}^d l }{(k^2+2kq)^{n_1}
    (l^2+2lq)^{n_2}(k^2)^{n_3}((k-l)^2-M^2)^{n_4}(l^2)^{n_5}}
  \,,
  \nonumber\\
  \lefteqn{Y^-_N(n_1,n_2,n_3,n_4,n_5) =}\nonumber\\&&
  \frac{e^{2\epsilon\gamma_E}}{(i\pi^{d/2})^2} 
  \int\frac{ {\rm d}^d k {\rm d}^d l }{(k^2+2kq)^{n_1}
    (l^2-2lq)^{n_2}((k-l)^2+2q(k-l))^{n_3}(k^2)^{n_4}(l^2-M^2)^{n_5}}
  \,,
  \nonumber\\
  \lefteqn{H^+_C(n_1,n_2,n_3,n_4,n_5) =}\nonumber\\&&
  \frac{e^{2\epsilon\gamma_E}}{(i\pi^{d/2})^2} 
  \int\frac{ {\rm d}^d k {\rm d}^d l }{(k^2+2kq)^{n_1}
    ((l+q)^2)^{n_2}(k^2)^{n_3}((k-l)^2-M^2)^{n_4}(l^2+m^2)^{n_5}}
  \,,
  \nonumber\\
  \lefteqn{W^-_C(n_1,n_2,n_3,n_4,n_5) =}\nonumber\\&&
  \frac{e^{2\epsilon\gamma_E}}{(i\pi^{d/2})^2} 
  \int\frac{ {\rm d}^d k {\rm d}^d l }{(k^2+2kq)^{n_1}
    ((l-q)^2)^{n_2}((k-l+q)^2)^{n_3}(k^2)^{n_4}(l^2-M^2)^{n_5}}
  \,,
  \label{eq::FDclasses}
\end{eqnarray}
where $d=4-2\epsilon$ is the space-time dimension.
The corresponding graphical representation can be found in
Fig.~\ref{fig::FDclasses}.
The integrals are defined in Minkowskian space and the 
$i\varepsilon$ prescription is understood.

Within {\tt FORM} we apply the projectors,
identify the external momentum with the top quark mass and 
decompose the numerators in terms of the denominators. This leads to
a large number of scalar integrals which differ from each other by the
power of the individual propagators.

A conventional way to reduce an arbitrary integral of a 
certain kind to a small set of so-called master integrals is based on
integration-by-parts~\cite{Chetyrkin:1981qh}, 
which provides relations between several
integrals of different complexity. 
The proper combination of these relations leads to new 
ones, so that the iteration of this procedure can be used for a
systematic reduction of an arbitrary 
integral to a small set of master integrals.

In Ref.~\cite{Laporta:2001dd} 
an algorithm has been formulated that performs automatically the
aforementioned reduction for a given set
of recurrence relations.
Currently several implementations of this algorithm exist.
However, to our knowledge, only one program is publicly
available, {\tt AIR}~\cite{Anastasiou:2004vj}. 
{\tt AIR} is written in {\tt MAPLE}, which certainly constitutes a
serious restriction for large-scale problems like, e.g., 
four-loop vacuum 
integrals~\cite{Schroder:2002re,Sturm_diss,Chetyrkin:2004fq,Schroder:2005db}.
Nevertheless, for the problem at hand {\tt AIR} is well suited to
perform the reduction.
It is straightforward to compose an interface which, for a given
class of Feynman diagrams (cf. Eq.~(\ref{eq::FDclasses})), produces the
corresponding integration-by-parts relations, passes them to {\tt AIR} 
and transforms the output into a table that can be read into {\tt FORM}.
These tables can be used to express
each integral occuring in our expression in terms of a few master
integrals.

At this point a comment concerning {\tt AIR} is in order. Our
experience with {\tt AIR} shows that there can be situations where the
set of master integrals is not minimal, although the complete set of
recurrence relations has been provided.
Consider, e.g., the integrals $Y^-_N$ with $M=0$. It is 
well-known that only three master are
needed for the computation of the two-loop pure QCD corrections to
$Z_m^{\rm OS}$. However, the naive application of {\tt AIR} leads to
four master integrals.
A straightforward inspection of the involved
integrals makes it possible to relate the additional master to the known ones.
The same is also true for our types of integrals.

For the diagrams where a neutral boson is exchanged one
requires altogether nine master integrals. Six of them either contain only
one mass scale, are vacuum diagrams, 
or consist of a product of two one-loop integrals. They read
\begin{align}
  H_1 &= H^+_N(1,1,0,0,0)\,,\qquad & 
  H_2 &= H^+_N(0,0,1,1,1)\,,\qquad &
  H_3 &= H^+_N(1,1,0,1,0)\,,\nonumber\\
  Y_1 &= Y^-_N(1,1,1,0,0)\,,\qquad &
  Y_2 &= Y^-_N(1,0,0,0,1)\,,\qquad &
  Y_3 &= Y^-_N(1,1,0,0,1)\,.
  \label{eq::master_neutral1}
\end{align}
Explicit analytic results are given in Appendix~\ref{app::ae_of_masters}.
The remaining three master integrals 
\begin{align}
  H_4 &= H^+_N(1,0,0,1,1)\,,\qquad &
  H_5 &= H^+_N(2,0,0,1,1)\,,\qquad &
  %  \nonumber\\
  Y_4 &= Y^-_N(1,1,1,0,1)\,,
  \label{eq::master_neutral}
\end{align}
are less trivial. Analytic expressions for $H_4$, $H_5$ and $Y_4$
can be found in Ref.~\cite{Jegerlehner:2003py}. More precisely one has
\begin{align}
  H_4 &\leftrightarrow J_{012}(1,1,1,m^2,M^2)\,,
  \nonumber\\
  H_5 &\leftrightarrow J_{012}(1,2,1,m^2,M^2)\,,
  \nonumber\\
  Y_4 &\leftrightarrow V_{mmmM}(1,1,1,1)\,,
\end{align}
where the integrals $J_{012}(1,1,1,m^2,M^2)$ and $J_{012}(1,2,1,m^2,M^2)$ 
are given in Eq.~(3.20) and $V_{mmmM}(1,1,1,1)$ 
in Eq.~(3.26) of Ref.~\cite{Jegerlehner:2003py}. 
We have checked all master integrals by 
considering their evaluation in an
asymptotic expansion around the three kinematical regions
$m\ll M$, $m\approx M$, and $m\gg M$. 
Altogether we computed up to 16
expansion terms and found
complete agreement with the results in the literature.
As can be deduced from the results of
Appendix~\ref{app::ae_of_masters}, where more details are provided,
the inclusion of about five
expansion terms in each region provides jointly a good approximation
over almost the whole range in $m/M$.

In the case of charged boson exchange one gets in addition two simple
master integrals
\begin{align}
  W_1 &= W^-_C(1,1,1,0,0)\,,\qquad & W_2 &= W^-_C(1,1,0,0,1)\,,
  \label{eq::master_charged1}
\end{align}
and five two-scale integrals
\begin{align}
  H_6 &= H^+_C(0,1,1,1,0)\,,\qquad &
  H_7 &= H^+_C(0,1,1,1,-1)\,,\qquad &
  H_8 &= H^+_C(1,1,0,1,0)\,,
  \nonumber \\
  W_3 &= W^-_C(0,-1,1,1,1)\,, \qquad &
  W_4 &= W^-_C(1,1,1,0,1) \,.
  \label{eq::master_charged}
\end{align}
As we will see in Sections~\ref{sec::Zm} and~\ref{sec::Z2}, for the
physical applications of this paper an expansion for $m\gg M$
of the integrals in Eq.~(\ref{eq::master_charged}) is sufficient to
obtain final results which in the physical region
are equivalent to the exact expressions.

%- }}}
%- {{{ On-shell mass renormalization constant:

\section{\label{sec::Zm}On-shell mass renormalization constant}

\indent

In this Section we discuss the results for the on-shell mass renormalization
counterterm. The QCD corrections up to three loops can be
found in
Refs.~\cite{Gray:1990yh,Fleischer:1998dw,Chetyrkin:1999ys,Chetyrkin:1999qi,Melnikov:2000qh}.
The one-loop electroweak  
and two-loop mixed corrections for light quarks can be found 
in Refs.~\cite{Hempfling:1994ar} and~\cite{Kniehl:2004hf}, respectively. 
In this case it is sufficient to evaluate the 
limiting behaviour for 
$m_q^2\ll M^2$ (where $M$ represents a boson mass).
The corrections of order $\alpha\alpha_s$ 
for the top quark have been considered in
Ref.~\cite{Jegerlehner:2003py,Faisst:2003px,Faisst:2004gn}.
In a recent paper~\cite{Martin:2005ch} the two-loop relation between a minimal
subtracted and the on-shell mass has been considered in a more
general framework. However, the masses of the vector bosons have been
neglected.

The relation between the bare mass, $m_t^0$, and the one defined in the 
$\overline{\rm MS}$ and on-shell scheme, $\overline{m}_t$ and $m_t$,
is given by
\begin{eqnarray}
  m_t^0 &=& Z_m^{\overline{\rm MS}} \overline{m}_t \,\,=\,\, 
  Z_m^{\rm OS} m_t\,. 
\end{eqnarray}

In order to discuss the result for the mass renormalization constant 
it is convenient to consider the finite ratio
\begin{eqnarray}
  z_m &=& \frac{ m_t }{ \overline{m}_t }
  \,\,=\,\,\frac{Z_m^{\overline{\rm MS}}}{Z_m^{\rm OS}}
  \,\,=\,\, 
  1 
  + \frac{\alpha_s}{\pi} C_F z_m^{\rm QCD}
  + \frac{\alpha}{\pi s_W^2} z_m^{\rm ew}
  + \frac{\alpha\alpha_s}{\pi^2 s_W^2} C_F z_m^{\rm mix}
  \,,
\end{eqnarray}
where $s_W\equiv \sin\theta_W$ is the sine of the Weinberg angle,
$C_F=(N_c^2-1)/(2N_c)$ with $N_c=3$ for SU(3) and
\begin{eqnarray}
  z_m^{\rm ew} &=& 
    z_m^{(1),\rm ew} \logmbar{\mu t}
  + z_m^{(0),\rm ew}
  \,,\nonumber\\
  z_m^{(0),\rm ew} &=& 
  z_m^{H,\rm ew}(\yhbar) + z_m^{W,\rm ew}(\ywbar) 
  + z_m^{Z,\rm ew}(\yzbar) + z_m^{A,\rm ew} + z_m^{\rm tad,\rm ew}
  \,,\nonumber\\
  z_m^{\rm mix} &=& 
  z_m^{(2),\rm mix} \logmbar{\mu t}^2
  + z_m^{(1),\rm mix} \logmbar{\mu t}
  + z_m^{(0),\rm mix}
  \,,\nonumber\\
  z_m^{(0),\rm mix} &=& 
  z_m^{H,\rm mix}(\yhbar) + z_m^{W,\rm mix}(\ywbar) + z_m^{Z,\rm mix}(\yzbar) 
  + z_m^{A,\rm mix} + z_m^{\rm tad,\rm mix}
  \,,
  \label{eq::zm_split}
\end{eqnarray}
with an analog separation for $z_m^{(1),\rm mix}$ and $z_m^{(2),\rm mix}$.
We furthermore introduce the notation
\begin{eqnarray}
  &&\yhbar\,\,=\frac{\overline{m}_t}{M_H}\,,\quad
  \ywbar\,\,=\frac{\overline{m}_t}{M_W}\,,\quad
  \yzbar\,\,=\frac{\overline{m}_t}{M_Z}
  \,,\nonumber\\\mbox{}
  &&\logmbar{H} \,\,= -\ln(\yhbar^2)\,,\quad
  \logmbar{W} \,\,= -\ln(\ywbar^2)\,,\quad
  \logmbar{Z} \,\,= -\ln(\yzbar^2) \,, \quad
  \logmbar{\mu t} \,\,= -\ln\left ( \frac{\mu^2}{\overline{m}_t^2} \right ) 
  \,.
  \nonumber\\
  \label{eq::zmdef}
\end{eqnarray}

The renormalization constant in the $\overline{\rm MS}$ scheme
is given by (see, e.g., Ref.~\cite{Jegerlehner:2003py})
\begin{eqnarray}
  Z_m^{\overline{\rm MS}} &=&
  1 - \frac{\alpha_s}{\pi}C_F\frac{3}{4\epsilon}
  + \frac{\alpha}{4\pi s_W^2}\frac{1}{\epsilon}
  \left(
    \frac{1}{4}
  + \frac{5}{4} a_t^2 s_W^2
  - \frac{3}{4} v_t^2 s_W^2
  - \frac{4}{3} s_W^2
  + \frac{3}{8} \frac{{\overline m_t^2}}{M_W^2}
  \right.
  \nonumber\\&&\mbox{}
  \left.
  - \frac{1}{4}
  - 3 a_t^2 s_W^2 \frac{M_Z^2}{M_H^2}
  - \frac{1}{2} a_t^2 s_W^2
  - \frac{3}{8} \frac{M_H^2}{M_W^2}
  - \frac{3}{2} \frac{M_W^2}{M_H^2}
  + N_c \frac{\overline{m}_t^4}{M_W^2 M_H^2}
  \right)
  \nonumber\\&&\mbox{}
  + \frac{\alpha\alpha_s}{4\pi^2 s_W^2} C_F
  \Bigg[
    \frac{1}{\epsilon^2}
    \left(
    - \frac{3}{16}
    - \frac{15}{16} a_t^2 s_W^2
    + \frac{9}{16} v_t^2 s_W^2
    + s_W^2
    - \frac{9}{16} \frac{{\overline m_t^2}}{M_W^2}
    \right)
    \nonumber\\&&\mbox{}
    + \frac{1}{\epsilon}
    \left(
     \frac{9}{32}
    + \frac{21}{32} a_t^2 s_W^2
    - \frac{3}{32} v_t^2 s_W^2
    - \frac{1}{6} s_W^2
    + \frac{3}{8} \frac{{\overline m_t^2}}{M_W^2}
    \right)
    + \frac{N_c}{2\epsilon} \frac{\overline{m}_t^4}{M_W^2 M_H^2}
    \nonumber\\&&\mbox{}
    +\frac{1}{\epsilon^2}
    \left(
     \frac{3}{16}
    + \frac{9}{4} a_t^2 s_W^2 \frac{M_Z^2}{M_H^2}
    + \frac{3}{8} a_t^2 s_W^2
    + \frac{9}{32} \frac{M_H^2}{M_W^2}
    + \frac{9}{8} \frac{M_W^2}{M_H^2}
    - \frac{9N_c}{4} \frac{\overline{m}_t^4}{M_W^2 M_H^2}
    \right)
    \Bigg]
  \,,
  \nonumber\\
  \label{eq::ZMS}
\end{eqnarray}
with $a_t=1/(2s_Wc_W)$, $v_t=(1/2- 4 s_W^2/3)/(2s_Wc_W)$ and
$c_W = \sqrt{1-s_W^2} = M_W/M_Z$.
For convenience, in Eq.~(\ref{eq::ZMS})
the contribution from the tadpole diagrams is displayed 
separately in the second (${\cal O}(\alpha)$) and last line
(${\cal O}(\alpha\alpha_s)$). Furthermore, all term proportional to
$N_c$ originate from tadpole diagrams.

At this point a comment concerning the various gauge
parameters is in order. In the electroweak sector we adopt Feynman
gauge for the $W$ and $Z$ boson, however, we 
allow for a general gauge parameter $\xi$ for QCD defined via the
gluon propagator
\begin{eqnarray}
  D_g^{\mu\nu}(q) &=& -i\,
  \frac{g^{\mu\nu}- \xi \frac{q^\mu q^\nu}{q^2}}
  {q^2+ i\varepsilon}
  \,.
\end{eqnarray}
On general grounds the on-shell mass and
also $z_m$ has to be independent of $\xi$
which serves as a welcome check for our calculation.

In Eq.~(\ref{eq::zm_split}) the two-loop expression for 
$z_m$ is split into
contributions induced by the Higgs, $W$ and $Z$ boson (including the 
corresponding Goldstone parts), the photon ($A$) and the tadpole
diagrams.
In the following we present analytical results for the individual
contributions. 
As we will
see in the discussion below it is sufficient to consider the
limit $y_W\to\infty$ and $y_Z\to\infty$ in order to obtain
agreement with the exact result below the percent level.
For this reason we show only the corresponding analytical
expressions.
Since the Higgs boson mass is still unknown we present 
the exact result, but also handy expansions  
in the three limits $y_H\to0$, $y_H\to1$ 
and $y_H\to\infty$. Furthermore, for completeness
also the tadpole result $z_m^{\rm tad}$ is listed,
which can be extracted
from Refs.~\cite{Faisst:2003px,Kniehl:2004hf,piclum_dipl}. 

Compact expressions for the Higgs boson contribution to
$z_m^{H, \rm ew}$ and $z_m^{H, \rm mix}$ expressed in terms of (known) master
integrals are given in Eqs.~(\ref{eq::zmew}) and ~(\ref{eq::zmmix})
in Appendix~\ref{app::zm},
where also the results for the $\mu$-dependent terms
$z_m^{(1),H,\rm mix}$ and $z_m^{(2),H,\rm mix}$ can be found.
The expansions in the physically interesting regions read
for the one-loop results

{\scalefont{0.8}
\begin{eqnarray}
  z_m^{\rm QCD} &=& 1
  \,,
  \nonumber\\
  z_{m,0}^{H, \rm ew} &=& \ywbar^2\Bigg[
     -\frac{5}{64} +   \frac{3\logmbar{H}}{32}
    +  \left( - \frac{1}{96} + \frac{\logmbar{H}}{16} \right) \yhbar^2 
    +  \left( - \frac{7}{128} + \frac{3\logmbar{H}}{32} \right) \yhbar^4 
    +  \left( - \frac{47}{320} + \frac{3\logmbar{H}}{16}\right) \yhbar^6
    \nonumber\\&&\mbox{}
    +  \left( - \frac{379}{960} + \frac{7\logmbar{H}}{16}\right) \yhbar^8 +{\mathcal{O}}(\yhbar^{10})
    \Bigg]
  \,,
  \nonumber\\
  z_{m,1}^{H, \rm ew} &=& \ywbar^2\Bigg[
    -\frac{3}{16} +   \frac{\pi \sqrtthree}{32}
  -  \frac{1}{16} \yhbarone  
  +  \left(\frac{1}{32} -  \frac{\pi \sqrtthree}{96}\right) \yhbarone^2 
  +  \left(\frac{1}{96} -  \frac{\pi \sqrtthree}{216}\right)\yhbarone^3
  \nonumber\\&&\mbox{}
  +  \left(\frac{1}{384} -  \frac{\pi \sqrtthree}{432}\right)\yhbarone^4
  +  \left( \frac{1}{960} - \frac{\pi \sqrtthree}{648} \right)\yhbarone^5
   + {\mathcal{O}}(\yhbarone^{6})  \Bigg]
  \,,
  \nonumber\\
  z_{m,\infty}^{H, \rm ew} &=& \ywbar^2\Bigg[
    -\!\frac{7}{32}
    +  \frac{\pi}{8}\frac{1}{\yhbar}
    +\!  \left(-\frac{3}{32} + \frac{3\logmbar{H}}{32}\right)\frac{1}{\yhbar^2}
    - \frac{3\pi}{64}\frac{1}{\yhbar^3} 
    +\!  \left(\frac{1}{24} - \frac{\logmbar{H}}{64}\right)\frac{1}{\yhbar^4}
    +  \frac{3\pi}{1024} \frac{1}{\yhbar^5}
\nonumber \\&&\mbox{}
    -  \frac{1}{640}\frac{1}{\yhbar^6} + {\mathcal{O}}\left (\frac{1}
  {\yhbar^{7}}\right )
    \Bigg]
  \,,
  \nonumber\\
  z_{m,\infty}^{W, \rm ew} &=& 
  \ywbar^2\Bigg[
    -\frac{1}{16}
    +  \left(-\frac{1}{16}  +  \frac{\logmbar{W}}{16}\right)\frac{1}{\ywbar^2}
    +  \left( \frac{3}{64} +  \frac{3\logmbar{W}}{32}\right) \frac{1}{\ywbar^4}
    +  \left( \frac{1}{12}  -  \frac{\logmbar{W}}{16} \right) \frac{1}{\ywbar^6}
    + {\mathcal{O}}\left (\frac{1}
  {\ywbar^{8}}\right )\Bigg]
  \,,
  \nonumber\\
  z_{m,\infty}^{Z, \rm ew} &=& \ywbar^2\Bigg[
     \frac{1}{32}
  +  \left(\frac{1}{32} +  \frac{\logmbar{Z}}{32}\right)\frac{1}{\yzbar^2}
  -  \frac{\pi}{32} \frac{1}{\yzbar^3}
  +  \left(\frac{1}{32} - \frac{\logmbar{Z}}{64}\right)\frac{1}{\yzbar^4}  
  +  \frac{\pi}{256}\frac{1}{\yzbar^5}
  -  \frac{1}{384}\frac{1}{\yzbar^6}
  + {\mathcal{O}}\left (\frac{1}
  {\yzbar^{7}}\right )\Bigg]
  \nonumber\\&&\mbox{}
 + a_t^2 s_W^2 \Bigg[
    -\frac{1}{2}
    + \frac{3\pi}{8}\frac{1}{\yzbar}
    + \left( - \frac{5}{16} + \frac{\logmbar{Z}}{4}\right)\frac{1}{\yzbar^2}
    - \frac{7\pi}{64}\frac{1}{\yzbar^3}
    + \left(  \frac{3}{32} -\frac{\logmbar{Z}}{32}\right)\frac{1}{\yzbar^4}
    + \frac{5\pi}{1024}\frac{1}{\yzbar^5}\nonumber \\&&\mbox{}
    - \frac{1}{480}\frac{1}{\yzbar^6} + {\mathcal{O}}\left (\frac{1}
  {\yzbar^{7}}\right )
    \Bigg ]
  + v_t^2 s_W^2 \Bigg[ \frac{1}{4}
    - \frac{\pi}{8}\frac{1}{\yzbar}
    + \frac{3}{16}\frac{1}{\yzbar^2}
    - \frac{3\pi}{64}\frac{1}{\yzbar^3}
    + \left( \frac{5}{96} - \frac{\logmbar{Z}}{32} \right)\frac{1}{\yzbar^4}
  \nonumber\\&&\mbox{}
    + \frac{9\pi}{1024}\frac{1}{\yzbar^5}
    - \frac{1}{160}\frac{1}{\yzbar^6} + {\mathcal{O}}\left (\frac{1}
  {\yzbar^{7}}\right )
    \Bigg] 
  \,,
  \nonumber\\
  z_{m}^{A, \rm ew} &=& \frac{4}{9}s_W^2
  \,,
  \nonumber\\
  z_{m}^{tad, \rm ew} &=& \ywbar^2\Bigg[
    \left(\frac{3}{32} -  \frac{3\logmbar{H}}{32}\right)\frac{1}{\yhbar^2}
    + \left(\frac{1}{8} 
    - \frac{3\logmbar{W} }{8}\right)\frac{\yhbar^2}{\ywbar^4}
    + \left(\frac{1}{16} 
    - \frac{\logmbar{W}}{16}\right)\frac{1}{\ywbar^2} \Bigg]
  \nonumber\\&&\mbox{}
  +  a_t^2 s_W^2  
  \Bigg [
    \left(\frac{1}{4} - \frac{3 \logmbar{Z}}{4}\right)\frac{\yhbar^2}{\yzbar^2}
    + \left(\frac{1}{8} -  \frac{\logmbar{Z}}{8}\right)
    \Bigg ]
  -\frac{N_c}{4} \ywbar^2 \yhbar^2
  \,,
  \label{eq::zmew_res}
\end{eqnarray}
}

\noindent
with $\yhbarone=1-1/\yhbar^2$.
The subscripts $0$, $1$ and $\infty$ indicate the cases
${\overline m_t}\ll M_H$, ${\overline m_t} \approx M_H$, and ${\overline
  m_t} \gg M_{H,W,Z}$, respectively.
The coefficients in front of $\logmbar{\mu t}$ are given by

{\scalefont{0.8}
\begin{align}
  &z_m^{(1),{\rm QCD}} = -\frac{3}{4} \, ,\quad\quad
  z_{m}^{(1),H,\rm ew} = \frac{3\ywbar^2}{32} \,,\quad\quad
  z_{m}^{(1),W,\rm ew} = \ywbar^2 \Bigg ( \frac{1}{32} +
  \frac{1}{16\ywbar^2} \Bigg ) \, ,
  \nonumber \\
  &z_{m}^{(1),Z,\rm ew} = -\frac{\ywbar^2}{32} + \frac{5 a_t^2 s_W^2}{16} -
  \frac{3 v_t^2 s_W^2}{16}\,, \quad\quad
  z_m^{(1),A,\rm ew} =  -\frac{s_W^2}{3} \, ,
  \nonumber\\
  &z_m^{(1),tad, \rm ew} = \ywbar^2 \Bigg ( -\frac{3}{32\yhbar^2} -
  \frac{3\yhbar^2}{8\ywbar^4} - \frac{1}{16\ywbar^2} \Bigg ) 
  + a_t^2 s_W^2 \Bigg (
  -\frac{3 \yhbar^2}{4\yzbar^2} - \frac{1}{8} \Bigg )
  +\frac{N_c}{4} \ywbar^2 \yhbar^2
  \, .
\end{align}
}

The corresponding two-loop expressions are slightly more lengthy but
still rather compact. They read

{\scalefont{0.8}
\begin{eqnarray}
  \lefteqn{  z_{m,0}^{H,\rm mix} = 
  \ywbar^2\Bigg[
    -\frac{9}{32} +   \frac{33\logmbar{H}}{128} -
    \frac{9\logmbar{H}^2}{128} -\frac{3\pi^2}{128} 
    +  \left(  - \frac{49}{1152} + \frac{185\logmbar{H}}{288} +
    \frac{13\logmbar{H}^2}{192} -\frac{\pi^2}{192}  \right) \yhbar^2  
  }\nonumber\\&&\mbox{}
    +  \left( - \frac{21097}{18432} + \frac{665\logmbar{H}}{768}  +
    \frac{45\logmbar{H}^2}{256} +\frac{15\pi^2}{256} \right) \yhbar^4  
    +  \left( - \frac{4145731}{1152000} + \frac{45727\logmbar{H}}{28800}  +
    \frac{881\logmbar{H}^2}{1920} +\frac{379\pi^2}{1920} \right) \yhbar^6
    \nonumber\\&&  \mbox{}
    +  \left( - \frac{4652609}{432000} + \frac{51199\logmbar{H}}{14400}  +
    \frac{313\logmbar{H}^2}{240} +\frac{49\pi^2}{80} \right) \yhbar^8  
    + {\mathcal{O}} (\yhbar^{10})\Bigg]
  \,,
  \nonumber\\ 
  \lefteqn{z_{m,1}^{H,\rm mix} =
  \ywbar^2 \Bigg\{
    -  \frac{133}{256}
    +  \frac{\nyeptwo}{32}
    +  \left ( \frac{\logthree}{48}
    - \frac{35}{768} \right ) \pi^2
    +  \left (\frac{9 \logthree}{32}
    + \frac{531}{256} \right ) S_2
    - \frac{\sqrtthree}{16} \lsthree
    - \frac{7\sqrtthree}{1152} \pi^3
    }\nonumber\\&&  \mbox{}
    +   \left (- \frac{\logthree}{24}
    - \frac{\logthreetwo}{192}
    +  \frac{3 S_2}{32}
    + \frac{53}{384} \right ) \pi \sqrtthree -\frac{\zeta(3)}{6}
    +  \left[ \frac{41}{96} + \frac{\nyeptwo}{24} + \left (
    \frac{\logthree}{36} 
    + \frac{1}{144} \right ) \pi^2    
    + \left ( \frac{3\logthree}{8}
    - \frac{93}{64} \right ) S_2  
    \right.\nonumber\\&&\left.  \mbox{}
    - \frac{\sqrtthree}{12} \lsthree
    -  \frac{7\sqrtthree}{864} \pi^3 
    +  \left (
       \frac{\logthree}{144}
    - \frac{\logthreetwo}{144} + \frac{S_2}{8}
    - \frac{13}{288} \right ) \pi \sqrtthree  - \frac{2\zeta(3)}{9} 
    \right] \yhbarone 
    \nonumber\\&&  \mbox{}
    +  \left [ \frac{1}{12} + \frac{\nyeptwo}{96} + \left ( 
    \frac{\logthree}{144} + \frac{7}{216} \right ) \pi^2 +
    \left ( \frac{3\logthree}{32} - \frac{123}{128} \right ) S_2
    - \frac{\sqrtthree}{48} \lsthree  - \frac{7\sqrtthree}{3456} \pi^3
    \right.\nonumber\\&&\left.  \mbox{}
    +  \left (   - \frac{\logthree}{288}
    - \frac{\logthreetwo}{576} +  \frac{S_2}{32} - \frac{1}{36} \right
    ) \pi \sqrtthree 
    - \frac{\zeta(3)}{18} \right ] \yhbarone^2
    +  \left [ -\frac{1}{288}  + \frac{143\pi^2}{7776}
    -  \frac{19 S_2}{64} 
    \right.\nonumber\\&&  \mbox{}\left.
    -  \left ( \frac{\logthree}{216}
    + \frac{59}{5184} \right ) \pi \sqrtthree 
    \right ] \yhbarone^3
    +  \left [ \frac{55}{6912} +  \frac{4145\pi^2}{373248}
    - \frac{257 S_2}{1536}
    -  \left ( \frac{5\logthree}{5184}
    + \frac{695}{62208} \right ) \pi \sqrtthree
    \right ] \yhbarone^4
    \nonumber\\&&  \mbox{}
    +  \left [ \frac{163}{34560}  + \frac{3923\pi^2}{466560} -
    \frac{71 S_2}{640} 
    - \left ( \frac{\logthree}{2160} + \frac{289}{31104} \right ) \pi
    \sqrtthree 
    \right ] \yhbarone^5 + {\mathcal{O}} ( \yhbarone^{6} )\Bigg\} 
  \, , \nonumber \\
  \lefteqn{z_{m,\infty}^{H,\rm mix} = \ywbar^2 \Bigg\{
    -\frac{187}{256}
    +  \left ( \frac{\logtwo}{4}
    - \frac{1}{64} \right ) \pi^2 - \frac{3\zeta(3)}{8}
    -  \frac{\pi}{8} \frac{1}{\yhbar}
    + \left[ \frac{39}{64} -  \frac{15\logmbar{H}}{64} - \frac{3\logtwo}{16} \pi^2  + \frac{9\zeta(3)}{32}
    \right ] \frac{1}{\yhbar^2}
    }\nonumber \\&& \mbox.{}
    +  \left [ \frac{173}{576} - \frac{\logmbar{H}}{24} - \frac{\logtwo}{6}
    \right ] \pi \frac{1}{\yhbar^3} +  \left [ - \frac{281}{1536} +
    \frac{17\logmbar{H}}{256} + \left ( \frac{\logtwo}{32} +
    \frac{3}{256} \right ) \pi^2  - \frac{3\zeta(3)}{64}
    \right] \frac{1}{\yhbar^4} \nonumber \\&& \mbox{}
    + \left [ - \frac{4861}{76800} +\frac{3\logmbar{H}}{320} +  \frac{3
      \logtwo}{80}  \right ] \pi \frac{1}{\yhbar^5} +  \left [
    \frac{85}{13824}  - \frac{5\pi^2}{3072}  \right ] 
    \frac{1}{\yhbar^6} + {\mathcal{O}}\left (\frac{1}{\yhbar^{7}}\right )
    \Bigg\}
  \, , \nonumber
\end{eqnarray}
\begin{eqnarray}
  \lefteqn{z_{m,\infty}^{W,\rm mix} =
    \ywbar^2 \Bigg[  -\frac{65}{256} +  \frac{5\pi^2}{128}
      -\frac{3\zeta(3)}{32} 
      -  \left ( \frac{19}{256} + \frac{\logmbar{W}}{32} \right
      )\frac{1}{\ywbar^2}  
      +  \left ( - \frac{15}{32} - \frac{15\logmbar{W}}{64} +  \frac{\pi^2}{64}
      + \frac{9\zeta(3)}{32} \right )\frac{1}{\ywbar^4}
    }\nonumber\\&&\mbox{}
    +  \left (- \frac{71}{384} + \frac{15\logmbar{W}}{64} - \frac{5\pi^2}{288}
    - \frac{3\zeta(3)}{16} \right )\frac{1}{\ywbar^6} +
    {\mathcal{O}}\left (\frac{1} 
    {\ywbar^{8}}\right ) \Bigg ]
  \,,\nonumber\\
  \lefteqn{z_{m,\infty}^{Z,\rm mix} =
    \ywbar^2\Bigg [ \frac{37}{256} - \frac{\pi^2}{64} + \left (
      -\frac{1}{64}-\frac{\logmbar{Z}}{64} + \left(-\frac{\logtwo}{16} +
      \frac{1}{32} \right ) \pi^2 + \frac{3\zeta(3)}{32}  \right )
    \frac{1}{\yzbar^2}  + \frac{\pi}{32}\frac{1}{\yzbar^3} 
    }\nonumber \\&& \mbox{} 
    + \left( -\frac{55}{512}+\frac{13\logmbar{Z}}{256}+
    \frac{\logtwo}{32}\pi^2 - \frac{3\zeta(3)}{64} \right)
    \frac{1}{\yzbar^4} + \left ( -\frac{667}{11520} + \frac{\logmbar{Z}}{96} +
    \frac{\logtwo}{24}  \right ) \pi \frac{1}{\yzbar^5} + \left(
    \frac{11}{1536}- \frac{3\pi^2}{1024}  \right ) \frac{1}{\yzbar^6}
    \nonumber \\&& \mbox{}
    + {\mathcal{O}}\left (\frac{1}{\yzbar^{7}}\right )\Bigg ]
  + a_t^2 s_W^2 \Bigg [ \frac{57}{256}+
    \left(\frac{3\logtwo}{4}-\frac{11}{32} \right ) \pi^2
    -\frac{9\zeta(3)}{8} -  \frac{3\pi}{8} \frac{1}{\yzbar} 
    \nonumber \\&& \mbox{} 
    + \left(
    \frac{7}{8} - \frac{13\logmbar{Z}}{16} + \left(-\frac{\logtwo}{2} +
    \frac{3}{32}  \right) \pi^2 + \frac{3\zeta(3)}{4} \right )
    \frac{1}{\yzbar^2}  
    + \left( \frac{167}{192} -\frac{\logmbar{Z}}{8} -\frac{\logtwo}{2}  \right
    ) \pi \frac{1}{\yzbar^3} 
    \nonumber \\&& \mbox{}
    + \left( -\frac{71}{256} +
    \frac{15\logmbar{Z}}{128} 
    + \left( \frac{\logtwo}{16} + \frac{5}{256} \right) \pi^2 -
    \frac{3\zeta(3)}{32}  \right ) \frac{1}{\yzbar^4} +
         {\mathcal{O}}\left (\frac{1} 
         {\yzbar^{5}}\right )\Bigg ] \nonumber \\&&
  \mbox{}
  + v_t^2 s_W^2 \Bigg [ -\frac{71}{256}+
    \left(-\frac{\logtwo}{4}+\frac{5}{32} \right ) \pi^2
    +\frac{3\zeta(3)}{8} + \left(
    - \frac{3}{4} + \frac{3\pi^2}{32} \right )
    \frac{1}{\yzbar^2} + \left( -\frac{19}{288} + \frac{\logmbar{Z}}{24} +
    \frac{\logtwo}{6}  \right ) \pi \frac{1}{\yzbar^3} \nonumber \\ &&
    \mbox{}  + \left( -\frac{39}{256} + \frac{15\logmbar{Z}}{128} + \left(
  \frac{\logtwo}{16} - \frac{7}{256} \right) \pi^2 -
  \frac{3\zeta(3)}{32}  \right ) \frac{1}{\yzbar^4} +
       {\mathcal{O}}\left (\frac{1} 
       {\yzbar^{5}}\right )\Bigg ]
  \,,\nonumber
\\
%\end{eqnarray}
%\begin{eqnarray}
  \lefteqn{z_{m}^{A,\rm mix} =
    \left[-\frac{71}{144}
    +  \left (  - \frac{4\logtwo}{9}+\frac{5}{18}  \right ) \pi^2 +
    \frac{2\zeta(3)}{3} \right] s_W^2
    \,,}\nonumber\\
  \lefteqn{z_{m}^{tad,{\rm mix}} =
    \ywbar^2 \Bigg [  \left ( -\frac{3}{64}
      + \frac{3\logmbar{H}}{64} \right ) \frac{1}{\yhbar^2}
      +  \left( - \frac{1}{16} + \frac{3}{16}
      \logmbar{W} \right )\frac{\yhbar^2}{\ywbar^4}
      +  \left ( - \frac{1}{32} + \frac{1}{32} \logmbar{W}  \right )
      \frac{1}{\ywbar^2} \Bigg ]
  }\nonumber\\&&\mbox{}
  +  a_t^2 s_W^2 \left [ 
  \left ( -\frac{1}{8}
  +  \frac{3\logmbar{Z}}{8} \right ) \frac{\yhbar^2}{\yzbar^2} +
  \left( -\frac{1}{16} 
  + \frac{\logmbar{Z}}{16} \right )  \right ]
  -\frac{N_c}{2} \ywbar^2 \yhbar^2
  \,,
  \label{eq::zm}
\end{eqnarray}
}

\noindent
where
\begin{eqnarray}
  \nyeptwo &=&  8 - \frac{\zeta(3)}{3} 
  + \frac{\pi^2}{6} +  \sqrtthree \left\{ -\frac{4\pi}{3} 
  -\frac{\pi^3}{36} + \frac{2\pi\logthree }{3} - \frac{\pi
    \logthreetwo}{6} 
  \right. \nonumber \\&& \left. 
  \mbox{} + 2 ( 2
  - \logthree) \left[  {\rm Ls_2}
  \left( \pi \right)  -  {\rm Ls_2} \left( \frac{2\pi}{3} \right )
  \right] + 2 \left[ {\rm Ls_3}
  \left( \pi \right )  -  {\rm Ls_3} \left( \frac{2\pi}{3} \right )
  \right] \right\}
  \nonumber \\
  &\approx&  0.245815004513
  \,,
  \nonumber\\
  S_2 &=& \frac{4\sqrtthree}{27} {\rm Ls}_2 \left( \frac{\pi}{3} \right) 
  \,\,\approx\,\,  0.260434137632
  \,,
  \nonumber\\
  \lsthree &\approx& -2.144767212569
  \,,
  \label{eq::const_def}
\end{eqnarray}
and $\zeta$ is Riemann's zeta function.
The definition of ${\rm Ls_i(z)}$ is given in Eq.~(\ref{eq::lsdef}).
The $\mu$ dependence is determined through

{\scalefont{0.8}
\begin{eqnarray}
  z_{m,0}^{(1),H,{\rm mix}} &=& \ywbar^2 \Bigg [ \frac{81}{256} -
    \frac{27\logmbar{H}}{128}  + \left( \frac{17}{128} -
    \frac{15\logmbar{H}}{64} \right ) \yhbar^2 + \left( \frac{219}{512} -
    \frac{63\logmbar{H}}{128} \right ) \yhbar^4 
    \nonumber\\&&\mbox{}
    + \left( \frac{1629}{1280} - \frac{81\logmbar{H}}{64} \right ) \yhbar^6
    + \left( \frac{5009}{1280} - \frac{231\logmbar{H}}{64} \right ) \yhbar^8
    + {\mathcal{O}} (\yhbar^{10})\Bigg ] \, , \nonumber \\
  z_{m,1}^{(1),H,{\rm mix}} &=& \ywbar^2 \Bigg [ \frac{33}{64} -
    \frac{9\pi\sqrtthree}{128} + \left( -\frac{3}{64} +
    \frac{\pi\sqrtthree}{32} \right ) \yhbarone + \left( -\frac{3}{128} +
    \frac{5\pi\sqrtthree}{384} \right ) \yhbarone^2 
    \nonumber \\&& 
    + \left( \frac{1}{128} +
    \frac{\pi\sqrtthree}{288} \right ) \yhbarone^3 
    + \left ( \frac{1}{512} +
    \frac{5\pi\sqrtthree}{1728} \right ) \yhbarone^4 + \left( \frac{3}{1280}
    + \frac{5 \pi \sqrtthree}{2592} \right ) \yhbarone^5 + {\mathcal{O}}(\yhbarone^{6} ) \Bigg ]  \, ,
    \nonumber \\ 
  z_{m,\infty}^{(1),H,{\rm mix}} &=& \ywbar^2 \Bigg [ \frac{63}{128} -
    \frac{3\pi}{16\yhbar} + \left( \frac{27}{128} - \frac{9 \logmbar{H}}{128}
    \right ) \frac{1}{\yhbar^2} + \left( \frac{1}{128} -
    \frac{3\logmbar{H}}{256} \right ) \frac{1}{\yhbar^4} +
    \frac{9\pi}{2048\yhbar^5} - \frac{9}{2560\yhbar^6}  \nonumber \\&&
  \mbox{} + {\mathcal{O}}\left (\frac{1}
  {\yhbar^{7}}\right )\Bigg ] 
  \, , \nonumber
%\\
\end{eqnarray}
\begin{eqnarray}
  z_{m,\infty}^{(1),W,{\rm mix}} &=& \ywbar^2 \Bigg [ \frac{3}{16} +
    \left ( \frac{5}{32} - \frac{3\logmbar{W}}{64} \right ) \frac{1}{\ywbar^2} +
    \left(\frac{45}{256} + \frac{9\logmbar{W}}{128} \right ) \frac{1}{\ywbar^4}
    + \left( \frac{3}{32} - \frac{9\logmbar{W}}{64} \right ) \frac{1}{\ywbar^6}
    + {\mathcal{O}}\left (\frac{1}{\ywbar^{8}}\right )\Bigg ]   \, , \nonumber \\
  z_{m,\infty}^{(1),Z,{\rm mix}} &=& \ywbar^2 \Bigg [ -\frac{15}{128} +
    \left( \frac{3}{128} - \frac{3\logmbar{Z}}{128} \right )
    \frac{1}{\yzbar^2} - \frac{3\logmbar{Z}}{256}
    \frac{1}{\yzbar^4} + 
    \frac{3\pi}{512 \yzbar^5} - \frac{3}{512 \yzbar^6}+ {\mathcal{O}}\left (\frac{1}
  {\yzbar^{7}}\right ) \Bigg ] \nonumber \\
  && + a_t^2 s_W^2 \Bigg [ \frac{5}{64} + \left( \frac{9}{64} +
    \frac{3\logmbar{Z}}{16} \right ) \frac{1}{\yzbar^2} - \frac{21\pi}{128\yzbar^3}
    + \left ( \frac{21}{128} - \frac{9\logmbar{Z}}{128} \right ) \frac{1}{\yzbar^4}
    + {\mathcal{O}}\left (\frac{1}{\yzbar^{5}}\right )\Bigg ] \nonumber \\
  && + v_t^2 s_W^2 \Bigg [ \frac{9}{64} + \frac{9}{64\yzbar^2} -
    \frac{9\pi}{128\yzbar^3} + 
    \left( \frac{9}{128} - \frac{9\logmbar{Z}}{128} 
    \right)\frac{1}{\yzbar^4} + {\mathcal{O}}\left (\frac{1}
  {\yzbar^{5}}\right )
    \Bigg ]   \, , \nonumber
%\\
\end{eqnarray}
\begin{eqnarray}
  z_{m}^{(1),A,{\rm mix}} &=& \frac{s_W^2}{4} \, , \nonumber \\
  z_{m}^{(1),tad,{\rm mix}} &=& \ywbar^2 \Bigg [ \left(
    -\frac{3}{128} + \frac{9\logmbar{H}}{128} \right ) \frac{1}{\yhbar^2} +
    \left( \frac{3}{32} + 
    \frac{9\logmbar{W}}{32} \right) \frac{\yhbar^2}{\ywbar^4} + \left(
  -\frac{1}{64} 
    + \frac{3\logmbar{W}}{64} \right ) \frac{1}{\ywbar^2} \Bigg ] \nonumber \\
  && + a_t^2 s_W^2 \Bigg [ \left( \frac{3}{16} + \frac{9\logmbar{Z}}{16}
    \right ) \frac{\yhbar^2}{\yzbar^2} - \frac{1}{32} +
  \frac{3\logmbar{Z}}{32} \Bigg]
  +\frac{11N_c}{16} \ywbar^2 \yhbar^2
\,,
\end{eqnarray}
\begin{align}
  &z_{m}^{(2),H,{\rm mix}} = -\frac{9\ywbar^2}{64} \,, \quad\quad
  z_{m}^{(2),W,{\rm mix}} = -\ywbar^2 \Bigg [ \frac{3}{64} +
    \frac{3}{64\ywbar^2} \Bigg ] 
  \,, \nonumber \\
  &z_{m}^{(2),Z,{\rm mix}} = \frac{3\ywbar^2}{64} - \frac{15 a_t^2
    s_W^2}{64} + \frac{9 v_t^2 s_W^2}{64} \,,\quad\quad
  z_{m}^{(2),A,{\rm mix}} = \frac{s_W^2}{4}
  \,, \nonumber \\
  &z_{m}^{(2),tad,{\rm mix}} = \ywbar^2 \Bigg [ \frac{9}{128\yhbar^2} +
    \frac{9\yhbar^2}{32\ywbar^4} + \frac{3}{64\ywbar^2} \Bigg ] +
  a_t^2 s_W^2 \Bigg [ 
    \frac{9\yhbar^2}{16\yzbar^2} + \frac{3}{32} \Bigg]
    -\frac{9N_c}{16} \ywbar^2 \yhbar^2\, .
\end{align}
}

\noindent
Note that the result for $z_{m}^{A}$ in the above expressions
can easily be extracted from the
two-loop pure QCD result~\cite{Gray:1990yh}.

\begin{figure}[t]
  \begin{center}
    \begin{tabular}{cc}
      \epsfig{figure=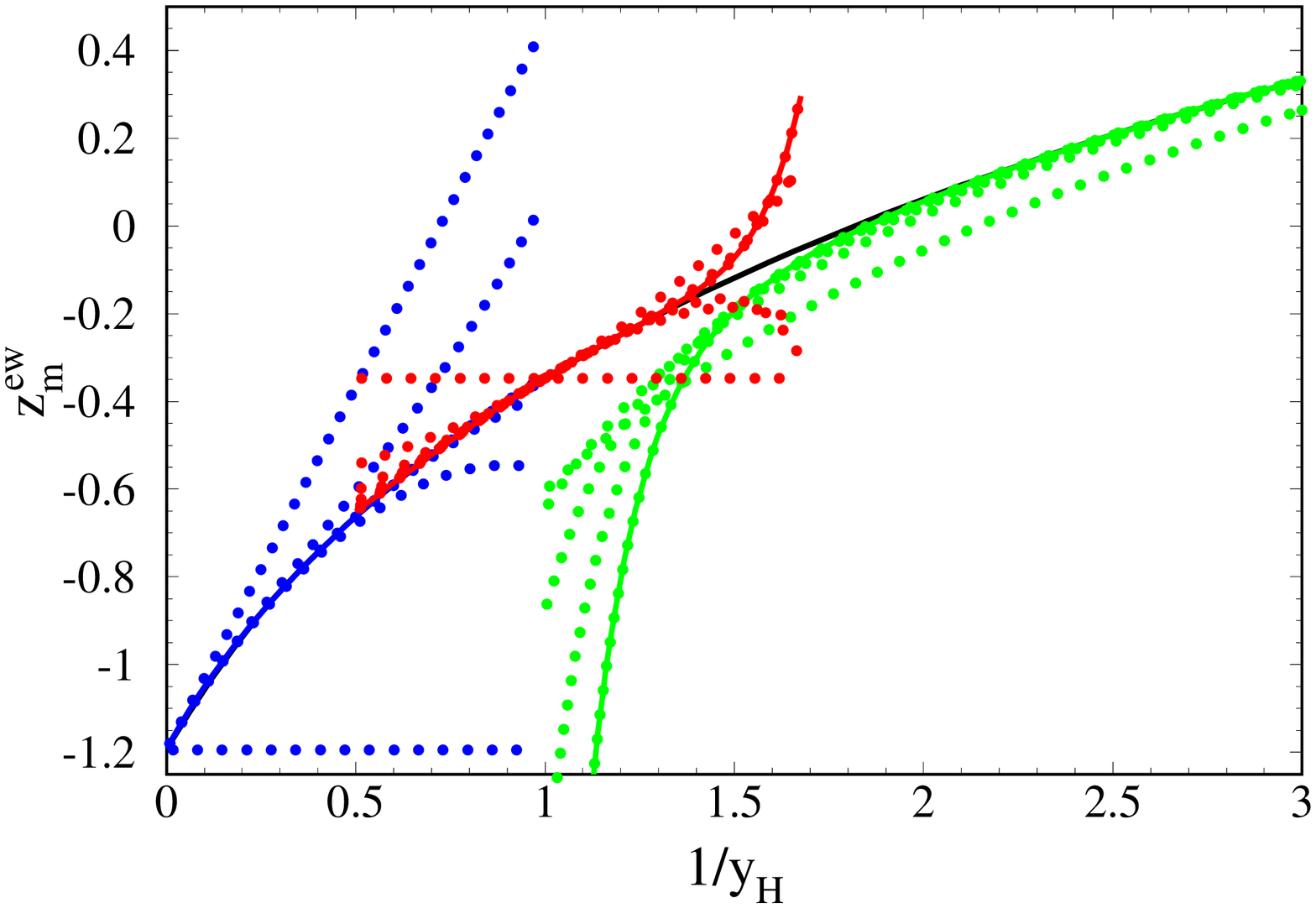,width=18em}
      &
      \epsfig{figure=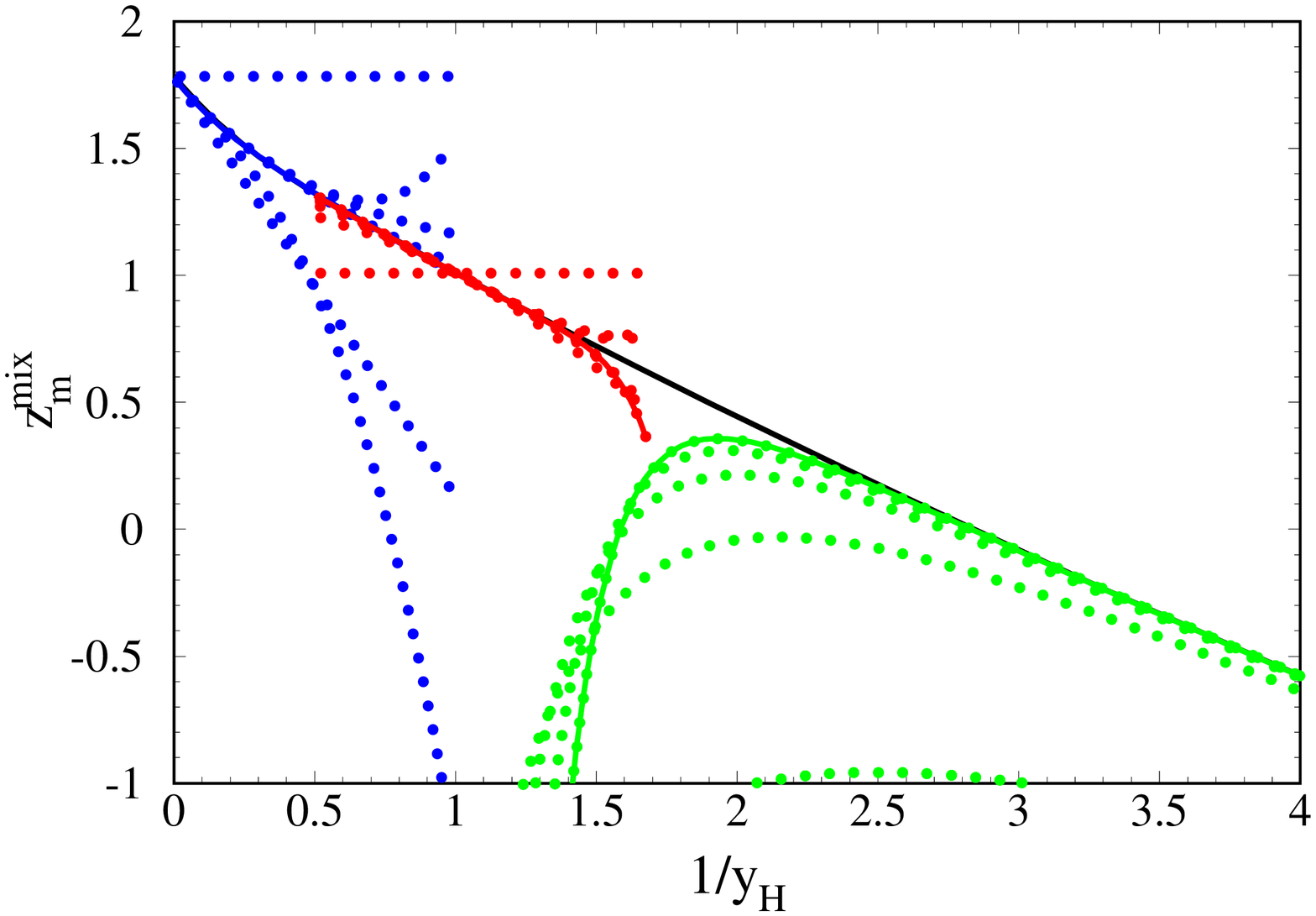,width=18em}
      \\(a) & (b)
    \end{tabular}
    \parbox{14.cm}{
      \caption[]{\label{fig::zm}\sloppy
        (a) One-loop and (b) two-loop corrections to $z_m$ as a
        function of $1/\yhbar=M_H/{\overline m_t}$. The solid (coloured) 
        lines correspond to the
        highest available order for each case. The dotted curves show 
        lower-order results and nicely demonstrate the convergence.
        The exact result, which is plotted (in black)
        over the whole $1/\yhbar$
        range, can be distinguished from the approximations only in 
        small gap around $1/\yhbar\approx1.5$.
        For the renormalization scale $\mu^2={\overline m_t^2}$ has
        been chosen. In the plots the contributions from the tadpole
        diagrams are not included.}} 
  \end{center}
\end{figure}

Let us next discuss the numerical consequences of our result and
compare the exact expressions with the compact expansions.
Actually the exact result from the $W$ and $Z$ boson contribution is 
very well reproduced both by the large-top mass limit or
the expansion around ${\overline m_t} \approx M_{W/Z}$. At the 
physical values for the particle masses we find a deviation from
the exact result for the $W$ and $Z$ contribution 
below the percent level,
which justifies the use of the expansion for the numerical evaluations.

The result for the one-loop coefficient of $\alpha/(\pi s_W^2)$
is shown in Fig.~\ref{fig::zm}(a) for $\mu={\overline m_t}$ as a function of 
$1/\yhbar=M_H/{\overline m_t}$
where the tadpole contributions~\cite{Faisst:2004gn},
which are numerically quite large,
have been subtracted for Feynman gauge.
Furthermore the following input values have been chosen
\begin{eqnarray}
  {\overline m_t} =165~\mbox{GeV}\,,\quad
  M_W=80.425~\mbox{GeV}\,,\quad
  M_Z=91.19~\mbox{GeV}\,,\quad
  c_W=M_W/M_Z\,.
\end{eqnarray}
Next to the (black) solid line which includes the exact result for the
Higgs mass dependence and the large-${\overline m_t}$ results for the 
$W$ and $Z$ contributions we also show the expansion terms
in the three kinematical regions (cf. Eq.~(\ref{eq::zmew_res})) as solid lines.
The lower-order results are plotted
as dotted lines in order to demonstrate the convergence of the 
approximations. It can be seen that
over almost the whole range of $\yhbar$ the expansion terms provide a very good
approximation to the exact result,
except for a small region with $\yhbar\approx 0.5\ldots 0.7$ which corresponds to
$M_H\approx 250 \ldots 300$~GeV.
We want to mention that in Fig.~\ref{fig::zm}(a) also the result of
Ref.~\cite{Jegerlehner:2003py} is shown which in contrast to ours 
also takes into account 
the exact dependence on $M_W$ and $M_Z$. No visible effect is observed.

In Fig.~\ref{fig::zm}(b) we show the two-loop coefficient of
$\alpha\alpha_s/(\pi^2 s_W^2)$ as a function of $1/\yhbar$, where again
the tadpole contribution of Eq.~(\ref{eq::zm}) is subtracted.
The result containing the exact $M_H$ dependence 
(cf. Eq.~(\ref{eq::zmmix})) is plotted together with the expansion results of
Eq.~(\ref{eq::zm}).
In addition to the highest expansion terms we show as dotted lines
also the lower-order ones. One can see how the approximations nicely
improve while including step-by-step the higher order terms.
Note that at two-loop order it is not possible to separate the result given
in Ref.~\cite{Jegerlehner:2003py} into the contributions from the
individual bosons and the tadpole contribution. 
Thus in order to check our results we again include the 
exact results from Ref.~\cite{Jegerlehner:2003py} in our plots
(after subtracting the tadpole contributions). Again
no difference is visible. In fact, 
adopting from Eq.~(\ref{eq::zm}) the large-${\overline m_t}$ terms we
obtain at the physical values of $M_W$ and $M_Z$ agreement with the 
exact result below the per cent level.
Equivalently we can use the expansion
around ${\overline m_t} \approx M_{W/Z}$ and get results with the same
level of accuracy.

The excellent description of the exact result by the expansion terms
together with the relative simplicity of the results in
Eq.~(\ref{eq::zm}) provides sufficient motivation
to apply in the next section the same approach to the 
wave function renormalization constant.
In particular this means that both exact results and expansions are 
considered for the Higgs boson contributions 
and approximate formulae for large top quark mass are derived for
the remaining parts.

%- }}}
%- {{{ On-shell wave function renormalization constant:

\section{\label{sec::Z2}On-shell wave function renormalization constants}

\indent

In this Section we consider the wave function renormalization 
constants for the top quark defined through Eq.~(\ref{eq::z2lr}).
Since $Z_2^{L,\rm OS}$ and $Z_2^{R,\rm OS}$ 
contain infra-red divergences which only cancel
when considering a physical quantity, it is not possible to form
a finite ratio analog to $z_m^{\rm OS}$ in Eq.~(\ref{eq::zmdef}).
Thus, in the following we consider the divergent contributions and
the finite parts separately. Furthermore, we switch to the
vector and axial-vector contribution using the formulae
\begin{eqnarray}
  Z_2^{V,\rm OS} &=& \frac{1}{2}
  \left( Z_2^{R,\rm OS} + Z_2^{L,\rm OS} \right)
  \,,\nonumber\\
  Z_2^{A,\rm OS} &=& \frac{1}{2}
  \left( Z_2^{R,\rm OS} - Z_2^{L,\rm OS} \right)
  \,,
\end{eqnarray}
where a non-zero contribution to $Z_2^{A,\rm OS}$ only arises from the 
$W$- and $Z$-boson.

In contrast to $Z_m^{\rm OS}$ the wave function renormalization needs
not to be gauge parameter independent since it does not pose a physical
quantity. As for the mass renormalization constant, we choose
Feynman gauge for the electroweak part but allow for arbitrary $\xi$
in the QCD sector. We observe that $\xi$ drops out in the case of
$Z_2^{V,\rm OS}$ up to the two-loop order,
which is in analogy to QCD where only the three-loop
result starts to depend on $\xi$~\cite{Melnikov:2000zc}.
On the other hand $Z_2^{A,\rm OS}$ is $\xi$ dependent starting from
order $\alpha\alpha_s$.

In order to present the results in a compact form we introduce the notation
\begin{eqnarray}
  Z_2^{X,\rm OS} &=& 1
  + \frac{\alpha_s}{\pi} C_F \delta Z_{2,X}^{\rm QCD}
  + \frac{\alpha}{\pi}
  \left(\frac{f_{2,X}^{\rm ew}}{\epsilon}
  + z_{2,X}^{\rm ew}  
  \right)
  + \frac{\alpha\alpha_s}{\pi^2} C_F \left(
  \frac{g_{2,X}^{\rm mix}}{\epsilon^2}
  + \frac{h_{2,X}^{\rm mix}}{\epsilon}
  + z_{2,X}^{\rm mix}
  \right)
  \,,\nonumber\\
\end{eqnarray}
with $X = V,A$ and the analog splitting as in Eq.~(\ref{eq::zm_split}).
In particular, the results for the individual coefficients
are split according to the dependence on the renormalization scale and
decomposed into contributions originating from
the Higgs, $W$ and $Z$ bosons and the 
photon\footnote{The tadpole diagrams do not
  contribute to the wave function renormalization.} 
where for $W$ and $Z$ again the large-$m_t$ limit is adopted.

In contrast to the quotient $z_m$, where it was found convenient to use
the ${\overline {\rm MS}}$-mass as expansion parameter, for
$Z_{2}^{X,\rm OS}$ the natural choice is to express the results in
terms of the on-shell mass. Thus, in analogy to 
Eq.~(\ref{eq::zmdef}) we introduce the following notation:
\begin{eqnarray}
  &&\yh\,\,=\frac{m_t}{M_H}\,,\quad
  \yw\,\,=\frac{m_t}{M_W}\,,\quad
  \yz\,\,=\frac{m_t}{M_Z}
  \,,\nonumber\\\mbox{}
  &&\logm{H} \,\,= -\ln(\yh^2)\,,\quad
  \logm{W} \,\,= -\ln(\yw^2)\,,\quad
  \logm{Z} \,\,= -\ln(\yz^2) \,, \quad
  \logm{\mu t} \,\,= -\ln\left ( \frac{\mu^2}{m_t^2} \right ) 
  \,.
  \nonumber\\
  \label{eq::z2def}
\end{eqnarray}
For the numerical value of the top quark mass we use $m_t=175$~GeV.

In the main text we again list the expansion terms and relegate the
exact expression for the Higgs boson contribution 
to the Appendix (cf. Eq.~(\ref{eq::z21lhexact})).
For the QCD result 
and the pole part of the one-loop electro-weak corrections
we obtain the following exact expressions
(see, e.g., Ref.~\cite{Denner:1991kt}) 
\begin{align}
  &\delta Z_2^{V, \rm QCD} = \left (-\frac{3}{4\epsilon} + \frac{3}{4} 
  \logm{\mu t} - 1 \right )\, , 
  && 
  \delta Z_2^{A, \rm QCD} = 0 \, ,  \nonumber\\
  &f_{2,V}^{H,\rm ew} = -\frac{\yw^2}{32} \, , 
  &&%  f_{2,A}^{H,\rm ew} =  0\, , 
  f_{2,V}^{A,\rm ew} = -\frac{s_W^2}{3}
  \,, \nonumber\\
  &f_{2,V}^{W,\rm ew} = -\frac{\yw^2}{32} -  \frac{1}{16}\, , &&
  f_{2,A}^{W,\rm ew} =  \frac{\yw^2}{32}  - \frac{1}{16}\, , \nonumber\\ 
  &f_{2,V}^{Z,\rm ew} = - \frac{\yw^2}{32} - \frac{s_W^2 \,a_t^2}{16} -
  \frac{s_W^2\,v_t^2}{16} \, , &&  f_{2,A}^{Z,\rm ew} = - \frac{1}{8}
  s_W^2\, a_t v_t \,.
  %f_{2,A}^{A,\rm ew} = 0\,.  
\end{align}
The finite one-loop contributions read

{\scalefont{0.8}
\begin{eqnarray}
  z_{2,V,0}^{H, \rm ew} &=& \yw^2\Bigg[
     -\frac{1}{64} + \frac{\logm{H}}{32}
    -  \frac{1}{16}  \yh^2 
    -  \left(  \frac{19}{128} - \frac{3\logm{H}}{32} \right) \yh^4 
    -  \left(  \frac{77}{160} - \frac{3\logm{H}}{8}\right) \yh^6
    \nonumber\\&&\mbox{}
    -  \left(  \frac{519}{320} - \frac{21\logm{H}}{16}\right) \yh^8
    + {\mathcal{O}} (\yh^{10} )\Bigg]
  \,,
  \nonumber\\
  z_{2,V,1}^{H, \rm ew} &=& \yw^2 \Bigg[
    \frac{1}{8} - \frac{\pi \sqrtthree}{32}
  + \left( \frac{1}{16} - \frac{ \pi \sqrtthree}{48}  \right ) \yhone  
  +  \left( - \frac{1}{8} +  \frac{5 \pi \sqrtthree}{288}\right) \yhone^2 
  +  \left( - \frac{7}{96} + \frac{\pi \sqrtthree}{108}\right)\yhone^3
  \nonumber\\&&\mbox{}
  +  \left( - \frac{5}{128} + \frac{5 \pi \sqrtthree}{1296}\right)\yhone^4
  +  \left( - \frac{7}{240} + \frac{5 \pi \sqrtthree}{1944} \right)\yhone^5
   + {\mathcal{O}} (\yhone^{6} ) \Bigg]
  \,,
  \nonumber\\
  z_{2,V,\infty}^{H,\rm ew} &=&
  \yw^2 \Bigg [ \left( \frac{7}{32} + \frac{\logm{H}}{8} \right) -
  \frac{3\pi}{16} \frac{1}{\yh} + \left ( \frac{3}{32} -
  \frac{3\logm{H}}{16}\right ) \frac{1}{\yh^2} +
  \frac{15\pi}{128} \frac{1}{\yh^3} \nonumber \\
&&  + \left( - \frac{7}{64} + \frac{3\logm{H}}{64} \right )
  \frac{1}{\yh^4} - \frac{21\pi}{2048} \frac{1}{\yh^5} + \frac{1}{160}
  \frac{1}{\yh^6} + {\mathcal{O}}\left (\frac{1}
  {\yh^{7}}\right )\Bigg  ]
  \,,\nonumber\\
  z_{2,V,\infty}^{W,\rm ew} &=& \yw^2 \Bigg [  -\left( \frac{1}{16} + \frac{\logm{W}}{16}
  \right )\frac{1}{\yw^2}  -
 \left( \frac{21}{64} + \frac{\logm{W}}{32} \right ) \frac{1}{\yw^4} +
  \left ( - \frac{5}{48}  + \frac{3\logm{W}}{16} \right ) \frac{1}{\yw^6}
  + \frac{47}{384} \frac{1}{\yw^8} 
  \nonumber\\&&\mbox{}
  + {\mathcal{O}}\left (\frac{1}
  {\yw^{10}}\right ) \Bigg ] 
  \,,
  \nonumber\\
  z_{2,V,\infty}^{Z,\rm ew} &=& 
  \yw^2 \Bigg [ - \frac{1}{32} - \left ( \frac{3}{32} +
  \frac{\logm{Z}}{16}\right ) \frac{1}{\yz^2} +
  \frac{5\pi}{64} \frac{1}{\yz^3} + \left( - \frac{5}{64} + \frac{3\logm{Z}}{64} \right )
  \frac{1}{\yz^4} - \frac{7\pi}{512} \frac{1}{\yz^5} + \frac{1}{96}
  \frac{1}{\yz^6} 
  \nonumber\\&& 
  + {\mathcal{O}}\left (\frac{1} {\yz^{7}}\right ) \Bigg ] 
  + a_t^2  s_W^2\Bigg [  \left( \frac{3}{4} + \frac{3 \logm{Z}}{8}  \right ) - \frac{9\pi}{16} \frac{1}{\yz} +\left( 
  \frac{3}{8} - \frac{\logm{Z}}{2} \right ) \frac{1}{\yz^2} + \frac{35\pi}{128}
  \frac{1}{\yz^3} 
  \nonumber \\&& 
  + \left( - \frac{1}{4} + \frac{3 \logm{Z}}{32} \right) \frac{1}{\yz^4} 
  - \frac{35\pi}{2048} \frac{1}{\yz^5} + \frac{1}{120} \frac{1}{\yz^6}
  + {\mathcal{O}}\left (\frac{1}
  {\yz^{7}}\right )\Bigg ] 
  + v_t^2 s_W^2 \Bigg [  - \left( \frac{1}{4}  + \frac{\logm{Z}}{8}
  \right ) 
  \nonumber\\&& 
  + \frac{3\pi}{16} \frac{1}{\yz} - 
  \frac{3}{8} \frac{1}{\yz^2} + \frac{15\pi}{128} \frac{1}{\yz^3} 
  +\left( - \frac{1}{8} + \frac{3 \logm{Z}}{32}  \right )
  \frac{1}{\yz^4} 
  - \frac{63\pi}{2048} \frac{1}{\yz^5} + \frac{1}{40} \frac{1}{\yz^6}
  + {\mathcal{O}}\left (\frac{1} {\yz^{7}}\right )\Bigg ] \,, 
  \nonumber \\
 z_{2,V}^{A,\rm ew} &=& -\frac{4 s_W^2}{9} \, , \nonumber \\
 % z_{2,A}^{H,\rm ew} &=& 0 \, , \nonumber \\
 z_{2,A,\infty}^{W,\rm ew} &=& \yw^2 \Bigg [  \frac{1}{16}  - \left(
   \frac{1}{16} + 
   \frac{\logm{W}}{16} \right ) \frac{1}{\yw^2}+
   \left( - \frac{3}{64} + \frac{5\logm{W}}{32} \right ) \frac{1}{\yw^4} +
   \left( \frac{5}{48}  - \frac{\logm{W}}{16} \right ) \frac{1}{\yw^6}
   - \frac{7}{384} 
   \frac{1}{\yw^8}  
   \nonumber\\&&\mbox{}
   + {\mathcal{O}}\left (\frac{1}
        {\yw^{10}}\right )  \Bigg ] \, , \nonumber \\ 
 z_{2,A,\infty}^{Z,\rm ew} &=& a_t v_t s_W^2 \Bigg [ - \frac{1}{4} +
   \frac{\pi}{4} \frac{1}{\yz} + 
   \left( - \frac{1}{8} + \frac{\logm{Z}}{4} \right )\frac{1}{\yz^2} -
   \frac{5\pi}{32} \frac{1}{\yz^3} + \left( \frac{7}{48}
   - \frac{\logm{Z}}{16} \right )  \frac{1}{\yz^4} + \frac{7\pi}{512}
   \frac{1}{\yz^5} 
   \nonumber\\&&\mbox{}
   - \frac{1}{120} \frac{1}{\yz^6}   +
        {\mathcal{O}}\left (\frac{1} 
        {\yz^{7}}\right ) \Bigg ] \, , 
 \nonumber \\
 % z_{2,A}^{A,\rm ew} &=& 0 \,,
\end{eqnarray}
}

\noindent
and the scale-dependent terms are given by
\begin{align}
  &
  z_{2,V}^{(1),H,{\rm ew}} = \frac{\yw^2}{32} \, , 
  &&
  z_{2,V}^{(1),A,{\rm ew}} = \frac{s_W^2}{3}  \, , 
  \nonumber \\
  &
  z_{2,V}^{(1),W,{\rm ew}} = \frac{\yw^2}{32} + \frac{1}{16} \, , 
  &&
  z_{2,A}^{(1),W,{\rm ew}} = - \frac{\yw^2}{32} + \frac{1}{16} \, , 
  \nonumber \\
  &
  z_{2,V}^{(1),Z,{\rm ew}} = \frac{\yw^2}{32} + \frac{a_t^2 s_W^2}{16} +
  \frac{v_t^2 s_W^2}{16} \, , 
  &&
  z_{2,A}^{(1),Z,{\rm ew}} =  \frac{a_t v_t s_W^2}{8}\,.
\end{align}

The double-pole parts of the two-loop result are still quite compact and can
be cast into the form
\begin{align}
  &g_{2,V}^{H,\rm mix}= \frac{3 \yw^2}{64}\, , 
  &&g_{2,V}^{A,\rm mix}= \frac{s_W^2}{4}\, , 
%  &&  g_{2,A}^{H,\rm mix}= 0\, ,
  \nonumber\\
  &g_{2,V}^{W,\rm mix}= \frac{3 \yw^2}{64} + \frac{3}{64}\, , 
  &&g_{2,A}^{W,\rm mix}= -\frac{(2+\xi) \yw^2}{128} - \frac{1-\xi}{64}
  \, , 
  \nonumber\\ 
  &g_{2,V}^{Z,\rm mix}= \frac{3 \yw^2}{64} + \frac{3}{64} s_W^2 a_t^2 +
  \frac{3}{64} s_W^2 v_t^2 \, , 
  &&  g_{2,A}^{Z,\rm mix}= -\frac{a_t v_t
    s_W^2 (1-\xi)}{32}  \, ,
%  &&  g_{2,A}^{A,\rm mix}= 0
\end{align}
whereas for the single poles we obtain the expansions\footnote{Note, that
  the exact expressions for the Higgs boson results are given in
  Eq.~(\ref{eq::z2Hpoles}).} 

{\scalefont{0.8}
\begin{eqnarray}
  h_{2,V,0}^{H, \rm mix} &=& \yw^2\Bigg[
     \frac{31}{256} - \frac{3\logm{H}}{128}
    +  \frac{3}{64}  \yh^2 
    +  \left( \frac{57}{512} - \frac{9\logm{H}}{128} \right) \yh^4 
    +  \left(  \frac{231}{640} - \frac{9\logm{H}}{32}\right) \yh^6
    \nonumber\\&&\mbox{}
    +  \left(  \frac{1557}{1280} - \frac{63\logm{H}}{64}\right) \yh^8
    + {\mathcal{O}} (\yh^{10} )\Bigg]
  \,,
  \nonumber\\
 h_{2,V,1}^{H,\rm mix} &=& \yw^2 \Bigg[
    \frac{1}{64} + \frac{3\pi \sqrtthree}{128}
  + \left( -  \frac{3}{64} + \frac{ \pi \sqrtthree}{64}  \right ) \yhone  
  +  \left(\frac{3}{32} - \frac{5 \pi \sqrtthree}{384}\right) \yhone^2 
  +  \left(\frac{7}{128} -  \frac{\pi \sqrtthree}{144}\right)\yhone^3
  \nonumber\\&&\mbox{}
  +  \left(\frac{15}{512} - \frac{5 \pi \sqrtthree}{1728}\right)\yhone^4
  +  \left( \frac{7}{320} - \frac{5 \pi \sqrtthree}{2592} \right)\yhone^5
    + {\mathcal{O}}(\yhone^{6})\Bigg] \, , \nonumber \\
  h_{2,V,\infty}^{H,\rm mix} &=& 
 \yw^2 \Bigg [ - \left( \frac{7}{128} + \frac{3\logm{H}}{32} \right) +
  \frac{9\pi}{64} \frac{1}{\yh} + \left ( -\frac{9}{128} +
  \frac{9\logm{H}}{64}\right ) \frac{1}{\yh^2} -
  \frac{45\pi}{512} \frac{1}{\yh^3} \nonumber \\
&&  + \left( \frac{21}{256} - \frac{9\logm{H}}{256} \right )
  \frac{1}{\yh^4} + \frac{63\pi}{8192} \frac{1}{\yh^5} 
  + {\mathcal{O}}\left (\frac{1}{\yh^{6}}\right )\Bigg  ]
\,,\nonumber
\end{eqnarray}
\begin{eqnarray}
  h_{2,V,\infty}^{W,\rm mix} &=& 
 \yw^2 \Bigg [ \frac{7}{64}+ \left( \frac{17}{128} + \frac{3\logm{W}}{64}
  \right )\frac{1}{\yw^2}  +
 \left( \frac{63}{256} + \frac{3\logm{W}}{128} \right ) \frac{1}{\yw^4} +
  \left ( \frac{5}{64}  - \frac{9\logm{W}}{64} \right )
  \frac{1}{\yw^6} + {\mathcal{O}}\left (\frac{1}
  {\yw^{8}}\right ) \Bigg ] \,,\nonumber\\
  h_{2,V,\infty}^{Z,\rm mix} &=& 
 \yw^2 \Bigg [ \frac{17}{128} + \left ( \frac{9}{128} +
  \frac{3\logm{Z}}{64}\right ) \frac{1}{\yz^2} -
  \frac{15\pi}{256} \frac{1}{\yz^3} + \left( \frac{15}{256} - \frac{9\logm{Z}}{256} \right )
  \frac{1}{\yz^4} + \frac{21\pi}{2048} \frac{1}{\yz^5} - \frac{1}{128}
  \frac{1}{\yz^6} \nonumber\\
  && + {\mathcal{O}}\left (\frac{1}
  {\yz^{7}}\right ) \Bigg ] + a_t^2  s_W^2\Bigg [ - \left( \frac{61}{128} + \frac{9
 \logm{Z}}{32} \right ) + \frac{27\pi}{64} \frac{1}{\yz} + \left( -
  \frac{9}{32} + \frac{3\logm{Z}}{8} \right ) \frac{1}{\yz^2} - \frac{105\pi}{512}
  \frac{1}{\yz^3} \nonumber \\
 && + \left(
  \frac{3}{16} - \frac{9 \logm{Z}}{128} \right) \frac{1}{\yz^4} + {\mathcal{O}}\left (\frac{1}
  {\yz^{5}}\right )\Bigg ] \nonumber
  \\ 
  && + v_t^2 s_W^2 \Bigg [  \left( \frac{35}{128}  + \frac{3\logm{Z}}{32} \right ) - \frac{9\pi}{64} \frac{1}{\yz} +
  \frac{9}{32} \frac{1}{\yz^2} - \frac{45\pi}{512} \frac{1}{\yz^3} +
 \left( \frac{3}{32} - \frac{9 \logm{Z}}{128}  \right ) \frac{1}{\yz^4} + {\mathcal{O}}\left (\frac{1}
  {\yz^{5}}\right ) \Bigg ]
\,,\nonumber
\end{eqnarray}
\begin{eqnarray}
  h_{2,V}^{A,\rm mix} &=& \frac{17 s_W^2}{24} \,,\nonumber\\
%  h_{2,A}^{H,\rm mix} &=& 0 \, , \nonumber \\
  h_{2,A,\infty}^{W,\rm mix} &=& \yw^2 \Bigg [ -\frac{ ( 3 + 2 \xi)}{64}
     + \left ( \frac{3 (-1 + 2\xi) }{128} - \frac{(1-\xi)\logm{W}}{64}
     \right ) \frac{1}{\yw^2} +\left ( - \frac{3 (1 - \xi) }{256} +  \frac{5 (1-\xi)\logm{W}}{128} \right ) \frac{1}{\yw^4}   \Bigg ] \, , \nonumber \\ 
  h_{2,A,\infty}^{Z,\rm mix} &=& a_t v_t s_W^2 \Bigg [
     \frac{(-5+8\xi)}{64} + \frac{(1-\xi)\pi}{16\yz} + \left (
     -\frac{1}{32} + \frac{\logm{Z}}{16} \right ) \frac{1-\xi}{\yz^2} -
     \frac{5(1-\xi)\pi}{128\yz^3} + \left( \frac{7}{192} -
     \frac{\logm{Z}}{64}  \right ) \frac{1-\xi}{\yz^4}   \Bigg ] 
  \,.
  \nonumber\\
%  h_{2,A}^{A,\rm mix} &=& 0 \, , 
\end{eqnarray}
}

The scale dependence is ruled by the following coefficients:
\begin{align}
  &
  h_{2,V}^{(1),H,{\rm mix}} = -\frac{3\yw^2}{32} \, ,
  &&
  h_{2,V}^{(1),A,{\rm mix}} = -\frac{s_W^2}{2} \, , 
  \nonumber \\
  &
  h_{2,V}^{(1),W,{\rm mix}} = -\frac{3\yw^2}{32} - \frac{3}{32} \, , 
  &&
  h_{2,A}^{(1),W,{\rm mix}} =
  \frac{(2+\xi)\yw^2}{64}+\frac{(1-\xi)}{32} \, , 
  \nonumber \\
  &
  h_{2,V}^{(1),Z,{\rm mix}} = -\frac{3\yw^2}{32} - \frac{3 a_t^2
    s_W^2}{32} - \frac{3 v_t^2 s_W^2}{32} \, ,
  &&
  h_{2,A}^{(1),Z,{\rm mix}} =  \frac{(1-\xi) a_t v_t  s_W^2}{16}\,.
\end{align} 

Let us in the following discuss the results for the finite
contribution to $Z_2^{\rm OS}$.
The result from the Higgs boson exchange expressed in terms of
master integrals is given in Eq.~(\ref{eq::z2mix})
of Appendix~\ref{app::z2}. The expansion terms
read

{{\scalefont{0.8}
\begin{eqnarray}
  \lefteqn{z_{2,V,0}^{H,{\rm mix}} =
  \yw^2\Bigg[
    \frac{119}{512} - \frac{29\logm{H}}{256} - \frac{3\logm{H}^2}{256}
    +  \left( \frac{509}{2304} + \frac{\logm{H}}{192} -
    \frac{3\logm{H}^2}{128} +\frac{13\pi^2}{384}  \right) \yh^2
    }\nonumber \\ && \mbox{}
    +  \left(\frac{7627}{9216} - \frac{13\logm{H}}{512}  -
    \frac{31\logm{H}^2}{128} -\frac{71\pi^2}{768} \right) \yh^4   +
    \left(\frac{739079}{256000} + \frac{39533\logm{H}}{57600}  - 
    \frac{2861\logm{H}^2}{3840} - \frac{1219\pi^2}{3840} \right) \yh^6 
    \nonumber\\&&\mbox{}
    + \left(\frac{1962407}{230400} + \frac{12253\logm{H}}{3840} - 
    \frac{305\logm{H}^2}{128} - \frac{23\pi^2}{24} \right) \yh^8
    + {\mathcal{O}} (\yh^{10} )\Bigg]
  \,,\nonumber
\end{eqnarray}
\begin{eqnarray}
  \lefteqn{z_{2,V,1}^{H,{\rm mix}} =
  \yw^2 \Bigg[ 
    -  \frac{79}{768}
    -  \frac{\nyeptwo}{24}
    +  \left ( -\frac{\logthree}{36}
    + \frac{245}{2304} \right ) \pi^2
    -  \left ( \frac{3 \logthree}{8}
    + \frac{519}{256} \right ) S_2
    + \frac{\sqrtthree}{12} \lsthree
    + \frac{7\sqrtthree}{864} \pi^3
    }\nonumber\\&&  \mbox{}
    +   \left ( \frac{73 \logthree}{1152}
    + \frac{\logthreetwo}{144}
    -  \frac{S_2}{8}
    - \frac{41}{1152} \right ) \pi \sqrtthree + \frac{2\zeta(3)}{9}
    \nonumber\\&&  \mbox{}
    +  \left ( -\frac{29}{32} - \frac{3\nyeptwo}{32} + \left (
  -\frac{\logthree}{16} 
    + \frac{79}{3456} \right ) \pi^2    + \left ( -\frac{27\logthree}{32}
    + \frac{531}{128} \right ) S_2  + \frac{3\sqrtthree}{16} \lsthree
    +  \frac{7\sqrtthree}{384} \pi^3 
    \right.\nonumber\\&&\left.  \mbox{}
    +  \left (
    -  \frac{7\logthree}{192}
    + \frac{\logthreetwo}{64} - \frac{9 S_2}{32}
    + \frac{47}{576} \right ) \pi \sqrtthree  + \frac{\zeta(3)}{2} 
    \right) \yhone 
    \nonumber\\&&  \mbox{}
    +  \left ( - \frac{7}{192} - \frac{\nyeptwo}{32} - \left ( 
    \frac{\logthree}{48} + \frac{1417}{20736} \right ) \pi^2 +
    \left ( -\frac{9\logthree}{32} + \frac{669}{256} \right ) S_2
    + \frac{\sqrtthree}{16} \lsthree  + \frac{7\sqrtthree}{1152} \pi^3
    \right.\nonumber\\&&\left.  \mbox{}
    +  \left (   \frac{19\logthree}{1152}
    + \frac{\logthreetwo}{192} -  \frac{3 S_2}{32} + \frac{139}{3456} \right
    ) \pi \sqrtthree 
    + \frac{\zeta(3)}{6} \right ) \yhone^2  \nonumber \\ &&
    \mbox{}
    +  \left ( \frac{295}{3456}  - \frac{7867\pi^2}{186624}
    +  \frac{865 S_2}{768} + \left ( \frac{55\logthree}{2592}
    + \frac{113}{15552} \right ) \pi \sqrtthree 
    \right ) \yhone^3
    \nonumber\\&&  \mbox{}
    +  \left ( \frac{277}{6912} -  \frac{4367\pi^2}{186624}
    + \frac{631 S_2}{768}
    +  \left ( \frac{29\logthree}{5184}
    + \frac{71}{7776} \right ) \pi \sqrtthree
    \right ) \yhone^4
    \nonumber\\&& \mbox{}
    +  \left ( \frac{4009}{207360}  - \frac{98903\pi^2}{5598720} +
  \frac{15463 S_2}{23040} 
    + \left ( \frac{253\logthree}{77760} + \frac{4447}{466560} \right )
  \pi \sqrtthree 
    \right ) \yhone^5 + {\mathcal{O}} (\yhone^{6})\Bigg ]
  \,,\nonumber\\
  \lefteqn{z_{2,V,\infty}^{H,{\rm mix}} =
  \yw^2 \Bigg [
    \frac{123}{256} - \frac{\logm{H}}{2} + \frac{3\logm{H}^2}{64} 
    +  \left (  \frac{15}{128}  - \frac{\logtwo}{2}
    \right ) \pi^2 + \frac{3\zeta(3)}{4}
    + \left( \frac{15}{16} - \frac{9 \logm{H}}{64} -
  \frac{9\logtwo}{32} \right ) \pi \frac{1}{\yh} 
    }\nonumber \\&& \mbox{}
    + \left( -\frac{1585}{1152} + \frac{329\logm{H}}{384} -
    \frac{11\logm{H}^2}{128} + \left( -\frac{7}{192}+
    \frac{15\logtwo}{32} \right ) \pi^2  - \frac{45\zeta(3)}{64} 
    \right ) \frac{1}{\yh^2} +  \left ( -\frac{2053}{2304} +
  \frac{311\logm{H}}{1536} 
  \right.\nonumber \\&& \mbox{}\left.
  + \frac{487\logtwo}{768} 
    \right ) \pi \frac{1}{\yh^3}  
    +  \left (  \frac{29029}{115200} -
    \frac{271\logm{H}}{1920} + \frac{11\logm{H}^2}{512} - \left (  
    \frac{13}{384} +\frac{3\logtwo}{32} \right ) \pi^2  + \frac{9\zeta(3)}{64}
    \right ) \frac{1}{\yh^4}
    \nonumber\\&&
    + \left (  \frac{28439}{153600}
    - \frac{1563\logm{H}}{40960} -  \frac{2811 
      \logtwo}{20480}  \right ) \pi \frac{1}{\yh^5}  + {\mathcal{O}}\left (\frac{1}{\yh^{6}}\right ) \Bigg ]
  \,,\nonumber\\
%\end{eqnarray}
%\begin{eqnarray}
  \lefteqn{z_{2,V,\infty}^{W,{\rm mix}} =
  \yw^2 \Bigg  [ \frac{79}{256} -  \frac{\pi^2}{64} + \frac{3\zeta(3)}{32}
    +  \left ( -\frac{3}{256} + \frac{7\logm{W}}{64} -
  \frac{3\logm{W}^2}{128} - \frac{\pi^2}{128} - \frac{3\zeta(3)}{16}
  \right )\frac{1}{\yw^2} 
  }\nonumber \\ && \mbox{}
+  \left ( \frac{661}{512} - \frac{\logm{W}}{256}-
  \frac{3\logm{W}^2}{256} +  \frac{5\pi^2}{96}
   - \frac{15\zeta(3)}{32} \right )\frac{1}{\yw^4}
+  \left ( - \frac{317}{768} + \frac{\logm{W}}{32} +
  \frac{9\logm{W}^2}{128} + \frac{79\pi^2}{576}
  + \frac{9\zeta(3)}{16} \right )\frac{1}{\yw^6} 
  \nonumber \\ && \mbox{}
  + {\mathcal{O}}\left (\frac{1}
  {\yw^{8}}\right )\Bigg ]
  \,,\nonumber
\end{eqnarray}
\begin{eqnarray}
  \lefteqn{z_{2,V,\infty}^{Z,{\rm mix}} =
    \yw^2\Bigg [ \frac{107}{256} - \frac{\pi^2}{128} + \left 
      (\frac{19}{128}+\frac{17\logm{Z}}{128} - \frac{3\logm{Z}^2}{128} +
      \left( 
      - \frac{5}{64}+\frac{5\logtwo}{32} \right ) \pi^2 -
      \frac{15\zeta(3)}{64}  \right ) 
      \frac{1}{\yz^2}  
    }\nonumber \\&&\mbox{} 
    + \left( -\frac{3}{16} + \frac{15\logm{Z}}{256} +
    \frac{15\logtwo}{128} \right ) \pi \frac{1}{\yz^3} 
    + \left( \frac{757}{4608}-\frac{11\logm{Z}}{96}+
    \frac{11 \logm{Z}^2}{512}+ 
    \left( \frac{1}{768} - \frac{3\logtwo}{32}  \right )\pi^2 +
    \frac{9\zeta(3)}{64} \right) 
    \frac{1}{\yz^4} 
    \nonumber \\&&\mbox{}
    + \left ( \frac{7837}{46080} - \frac{271\logm{Z}}{6144} -
    \frac{479 \logtwo}{3072}  \right ) \pi \frac{1}{\yz^5} 
    + \left(
    \frac{1091}{76800} + \frac{\logm{Z}}{1280} + \frac{21\pi^2}{2048}
    \right ) \frac{1}{\yz^6}+ {\mathcal{O}}\left (\frac{1} 
         {\yz^{7}}\right )
         \Bigg ] \nonumber \\&& \mbox{}
  + a_t^2 s_W^2 \Bigg [ -\frac{815}{256}-\frac{9\logm{Z}}{16} +
    \frac{9\logm{Z}^2}{64} 
    + \left(\frac{125}{128} - \frac{3\logtwo}{2}\right ) \pi^2
    + \frac{9\zeta(3)}{4} +  \left(\frac{27}{32}
    -\frac{27\logm{Z}}{64}  - \frac{27\logtwo}{32}  \right )\pi
    \frac{1}{\yz} \nonumber \\ && \mbox{} + \left(
    \frac{259}{288} + \frac{67\logm{Z}}{96} - \frac{\logm{Z}^2}{8} +
    \left( -\frac{59}{192} + \frac{5\logtwo}{4}  \right) \pi^2 -
    \frac{15\zeta(3)}{8} \right ) 
    \frac{1}{\yz^2} + \left( - \frac{337}{192} + \frac{281\logm{Z}}{512}
    +\frac{457\logtwo}{256}  \right 
    ) \pi \frac{1}{\yz^3} \nonumber \\&& \mbox{}
    + \left( -\frac{34577}{57600} +
    \frac{19\logm{Z}}{480} + \frac{11\logm{Z}^2}{256}
    - \left( \frac{43}{768} +\frac{3\logtwo}{16}\right) \pi^2 +
    \frac{9\zeta(3)}{32}  \right ) \frac{1}{\yz^4}+ {\mathcal{O}}\left
    (\frac{1} 
         {\yz^{5}}\right ) \Bigg ] \nonumber \\&&
  \mbox{}
  + v_t^2 s_W^2 \Bigg [ \frac{369}{256}+ \frac{\logm{Z}}{8} -
    \frac{3\logm{Z}^2}{64} +
    \left(- \frac{51}{128}+\frac{\logtwo}{2} \right ) \pi^2
    - \frac{3\zeta(3)}{4} + \left( - \frac{3}{16}+ \frac{9\logm{Z}}{64}
    + \frac{9\logtwo}{32}\right ) \pi \frac{1}{\yz} 
    \nonumber \\ && \mbox{} 
    + \left( 
    \frac{217}{144} - \frac{19\logm{Z}}{96} + \frac{\logm{Z}^2}{32} -
    \frac{49\pi^2}{192} \right ) 
    \frac{1}{\yz^2} 
    + \left( \frac{991}{2304} -
    \frac{41\logm{Z}}{1536} - 
    \frac{217\logtwo}{768}  \right ) \pi \frac{1}{\yz^3} 
    \nonumber \\ && \mbox{} 
    + \left(
    -\frac{17873}{57600} + \frac{\logm{Z}}{480} +
    \frac{11\logm{Z}^2}{256} + \left( \frac{65}{768} -
    \frac{3\logtwo}{16} \right) \pi^2 + \frac{9\zeta(3)}{32}  \right )
    \frac{1}{\yz^4} + {\mathcal{O}}\left (\frac{1}
         {\yz^{5}}\right )\Bigg ] 
  \,,\nonumber\\
  \lefteqn{z_{2,V}^{A,{\rm mix}} =  \left[\frac{433}{144}
    +  \left ( \frac{8\logtwo}{9}-\frac{49}{72}  \right ) \pi^2 -
    \frac{4\zeta(3)}{3} \right] s_W^2
    \,,}\nonumber
\end{eqnarray}
\begin{eqnarray}
  \lefteqn{z_{2,A,\infty}^{W,{\rm mix}} =  \yw^2 \Bigg  [ - \frac{31}{256} -
  \frac{3\xi}{32}  -  \frac{\pi^2}{32} + \frac{\xi\pi^2}{256}
  + \frac{3\zeta(3)}{32} 
  }\nonumber \\ && \mbox{}
  +  \left ( \frac{13}{256} + \frac{9\xi}{64} -
    \frac{3\logm{W}}{64} + \frac{3\xi\logm{W}}{64} +
  \frac{\logm{W}^2}{128} - \frac{\xi \logm{W}^2}{128}  +
  \frac{13\pi^2}{384} - \frac{7 \xi \pi^2}{384}  - \frac{3\zeta(3)}{8}
  \right )\frac{1}{\yw^2} 
  \nonumber \\ && \mbox{}
+  \left ( - \frac{129}{512} - \frac{3\xi}{512}  + \frac{21\logm{W}}{256} -
  \frac{21\xi\logm{W}}{256} - \frac{5\logm{W}^2}{256} + \frac{5 \xi
    \logm{W}^2}{256} +  \frac{11\pi^2}{192} + \frac{5\xi\pi^2}{192}
   + \frac{15\zeta(3)}{32} \right )\frac{1}{\yw^4} \nonumber \\ && \mbox{}
+  \left ( \frac{193}{2304} - \frac{29\xi}{1152} - \frac{5\logm{W}}{96}
  + \frac{\xi\logm{W}}{48} +
  \frac{\logm{W}^2}{128} - \frac{\xi\logm{W}^2}{128} - \frac{11\pi^2}{144}
  - \frac{\pi^2 \xi}{96} 
  - \frac{3\zeta(3)}{16} \right )\frac{1}{\yw^6} + {\mathcal{O}}\left (\frac{1}
  {\yw^{8}}\right )\Bigg ]
  \,,\nonumber\\
  \lefteqn{z_{2,A,\infty}^{Z,{\rm mix}} = a_t v_t s_W^2 \Bigg
  [\frac{9}{128}+\frac{3\xi}{8} - \left(\frac{37}{192} -
    \frac{\logtwo}{4}\right ) \pi^2 + \frac{\pi^2\xi}{192}
 - \frac{3\zeta(3)}{8} +  \left( \frac{3}{16}
  - \frac{\logm{Z}}{16} - \frac{\logtwo}{8}  \right )\pi
  \frac{1-\xi}{\yz}
  }\nonumber \\ && \mbox{} + \left(
   \frac{1}{8} - \frac{3\xi}{16} - \frac{3\logm{Z}}{32} - \frac{3 \xi
     \logm{Z}}{32} - \frac{(1-\xi) \logm{Z}^2}{32} +
  \left( \frac{5}{32} - \frac{\logtwo}{2}  \right) \pi^2 +
  \frac{3\zeta(3)}{4} \right ) 
 \frac{1}{\yz^2}  
   \nonumber \\&& \mbox{}
 + \left( \frac{595}{1152} + \frac{13\xi}{128} -
   \frac{49\logm{Z}}{384} - \frac{5\xi\logm{Z}}{128} -
   \frac{113\logtwo}{192} - \frac{5\xi\logtwo}{64} \right
) \pi \frac{1}{\yz^3} 
   \nonumber \\&& \mbox{}
 + \left( - \frac{1}{32} + \frac{31\xi}{576} +
  \frac{5\logm{Z}}{192} + \frac{\xi \logm{Z}}{96} - \frac{(1+\xi)\logm{Z}^2}{128}
  + \left( \frac{17}{768} + \frac{\logtwo}{8}\right) \pi^2 -
  \frac{3\zeta(3)}{16}  \right ) \frac{1}{\yz^4} + {\mathcal{O}}\left (\frac{1}
  {\yz^{5}}\right )\Bigg ]
  \,.
  \label{eq::z2res}
\end{eqnarray}
}

The results covering the scale dependence are given by

{{\scalefont{.8}
\begin{eqnarray}
  z_{2,V,0}^{(1),H,{\rm mix}} &=& \yw^2 \Bigg [ -\frac{31}{128} +
    \frac{3\logm{H}}{64} -\frac{3\yh^2}{32} + \left( - \frac{57}{256} +
    \frac{9\logm{H}}{64} \right ) \yh^4 
    + \left( -\frac{231}{320} + \frac{9\logm{H}}{16} \right ) \yh^6 
    \nonumber\\&&\mbox{}
    + \left( -\frac{1557}{640} + \frac{63\logm{H}}{32} \right ) \yh^8 
  + {\mathcal{O}}(\yh^{10} )\Bigg ] \, , \nonumber \\
z_{2,V,1}^{(1),H,{\rm mix}} &=& \yw^2 \Bigg [ - \frac{1}{32} -
  \frac{3\pi\sqrtthree}{64} + \left( \frac{3}{32} -
  \frac{\pi\sqrtthree}{32} \right ) \yhone + \left( -\frac{3}{16} +
  \frac{5\pi\sqrtthree}{192} \right ) \yhone^2 
  \nonumber \\ && 
  + \left( -\frac{7}{64} + \frac{\pi \sqrtthree}{72} \right ) \yhone^3 
  + \left( -\frac{15}{256} 
  + \frac{5\pi\sqrtthree}{864} \right ) \yhone^4 + \left (
  -\frac{7}{160} + \frac{5\pi\sqrtthree}{1296} \right ) \yhone^5 + {\mathcal{O}}(\yhone^{6} ) \Bigg ]
\, , \nonumber \\
z_{2,V,\infty}^{(1),H,{\rm mix}} &=& \yw^2 \Bigg [ \frac{7}{64} +
\frac{3\logm{H}}{16} - \frac{9\pi}{32\yh} + \left ( \frac{9}{64} -
  \frac{9\logm{H}}{32} \right ) \frac{1}{\yh^2} + \frac{45
  \pi}{256\yh^3} + \left ( -\frac{21}{128} + \frac{9\logm{H}}{128}
\right ) \frac{1}{\yh^4} 
\nonumber\\&&\mbox{}
- \frac{63\pi}{4096\yh^5} 
+ {\mathcal{O}}\left (\frac{1}
  {\yh^{6}}\right ) \Bigg ] \, , \nonumber \\ 
z_{2,V,\infty}^{(1),W,{\rm mix}} &=& \yw^2 \Bigg [ -\frac{7}{32} - \left
  ( \frac{17}{64} + \frac{3\logm{W}}{32} \right ) \frac{1}{\yw^2} -
\left( \frac{63}{128} + \frac{3\logm{W}}{64} \right ) \frac{1}{\yw^4} +
\left ( -\frac{5}{32} + \frac{9\logm{W}}{32} \right ) \frac{1}{\yw^6}
\nonumber\\&&\mbox{}
+ {\mathcal{O}}\left (\frac{1}
  {\yw^{8}}\right )\Bigg ] \, , \nonumber \\ 
z_{2,V,\infty}^{(1),Z,{\rm mix}} &=& \yw^2 \Bigg [ -\frac{17}{64} -
\left( \frac{9}{64} + \frac{3\logm{Z}}{32} \right ) \frac{1}{\yz^2} +
\frac{15\pi}{128\yz^3} + \left ( -\frac{15}{128} +
  \frac{9\logm{Z}}{128} \right ) \frac{1}{\yz^4} -
\frac{21\pi}{1024\yz^5} + \frac{1}{64\yz^6} 
\nonumber \\&& 
+ {\mathcal{O}}\left (\frac{1}  {\yz^{7}}\right )\Bigg ] 
+ a_t^2 s_W^2 \Bigg [ \frac{61}{64} + \frac{9\logm{Z}}{16} -
\frac{27\pi}{32\yz} + \left ( \frac{9}{16} - \frac{3\logm{Z}}{4} \right
) \frac{1}{\yz^2} + \frac{105\pi}{256\yz^3} 
\nonumber \\&& 
+ \left ( -\frac{3}{8} +
  \frac{9\logm{Z}}{64} \right ) \frac{1}{\yz^4} 
  + {\mathcal{O}}\left (\frac{1}
  {\yz^{5}}\right )\Bigg ] 
+ v_t^2 s_W^2 \Bigg [ -\frac{35}{64} - \frac{3\logm{Z}}{16} +
\frac{9\pi}{32\yz} - \frac{9}{16\yz^2} + \frac{45\pi}{256\yz^3}
\nonumber \\&& 
 +
\left ( -\frac{3}{16} + \frac{9\logm{Z}}{64} \right ) \frac{1}{\yz^4}
+ {\mathcal{O}}\left (\frac{1}
  {\yz^{5}}\right )\Bigg ] \, , \nonumber \\ 
z_{2,V}^{(1),A,{\rm mix}} &=& -\frac{17 s_W^2}{12} \, , \nonumber \\
z_{2,A}^{(1),W,{\rm mix}} &=& \yw^2 \Bigg [ \frac{3+ 2\xi}{32} + \left(
  \frac{3-6\xi}{64} + \frac{1-\xi}{32} \logm{W}  \right )
\frac{1}{\yw^2} + \left(  \frac{3}{128} - \frac{5\logm{W}}{64} \right
) \frac{1-\xi}{\yw^4} 
\nonumber\\&&\mbox{}
+ \left( - \frac{5}{96} + \frac{\logm{W}}{32}  \right )
\frac{1-\xi}{\yw^6} + {\mathcal{O}}\left (\frac{1}
  {\yw^{8}}\right )\Bigg ]\, , 
\nonumber \\ 
z_{2,A}^{(1),Z,{\rm mix}} &=& a_t v_t s_W^2 \Bigg [ \frac{5}{32} -
\frac{\xi}{4} -  \frac{(1-\xi) \pi}{8\yz} + \left( \frac{1}{16} -
  \frac{\logm{Z}}{8} \right ) \frac{1-\xi}{\yz^2} +
\frac{5(1-\xi)\pi}{64} 
\nonumber\\&&\mbox{}
+ \left( - \frac{7}{96} + \frac{\logm{Z}}{32}
\right) \frac{1-\xi}{\yz^4} + {\mathcal{O}}\left (\frac{1}
  {\yz^{5}}\right )\Bigg ]  \, , 
\nonumber
\end{eqnarray}
\begin{align}
  &z_{2,V}^{(2),H,{\rm mix}} &=& \frac{3\yw^2}{32} \, ,
  &&z_{2,V}^{(2),W,{\rm mix}} &=& \frac{3\yw^2}{32} + \frac{3}{32} \, ,
  \nonumber \\
  &z_{2,V}^{(2),Z,{\rm mix}} &=& \frac{3\yw^2}{32} + \frac{3 a_t^2
    s_W^2}{32} + \frac{3 v_t^2 s_W^2}{32} \, ,
  &&z_{2,V}^{(2),A,{\rm mix}} &=& \frac{s_W^2}{2} \, , 
  \nonumber \\
  &z_{2,A}^{(2),W,{\rm mix}} &=& - \frac{(2+\xi)\yw^2}{64} -
  \frac{1-\xi}{32} \, ,
  &&z_{2,A}^{(2),Z,{\rm mix}} &=& - \frac{(1-\xi) a_t v_t s_W^2}{16} \, .
\end{align}
}

\begin{figure}[t]
  \begin{center}
    \begin{tabular}{cc}
      \epsfig{figure=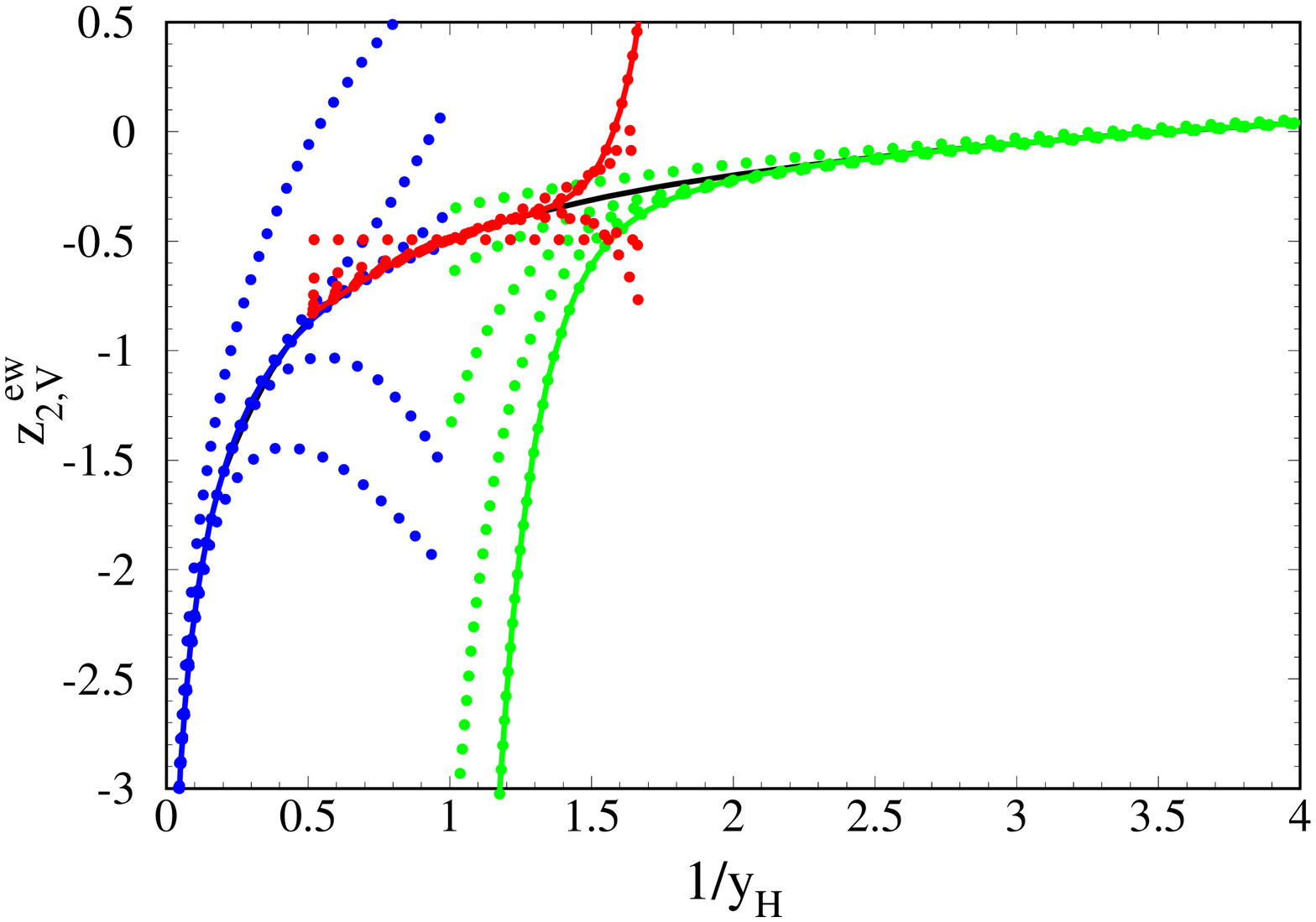,width=18em}
      &
      \epsfig{figure=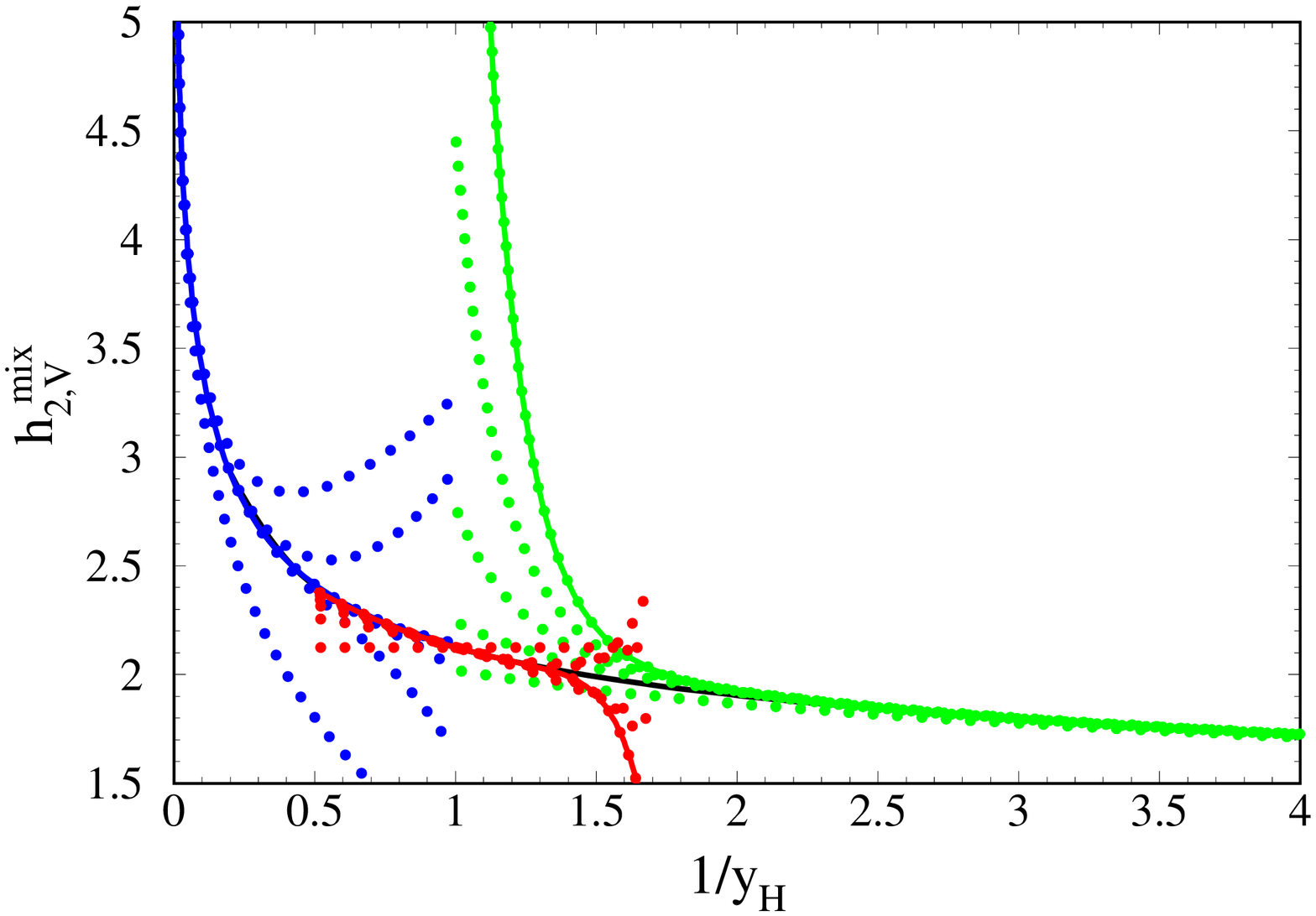,width=18em}
      \\(a) & (b)\\
      \multicolumn{2}{c}{
        \epsfig{figure=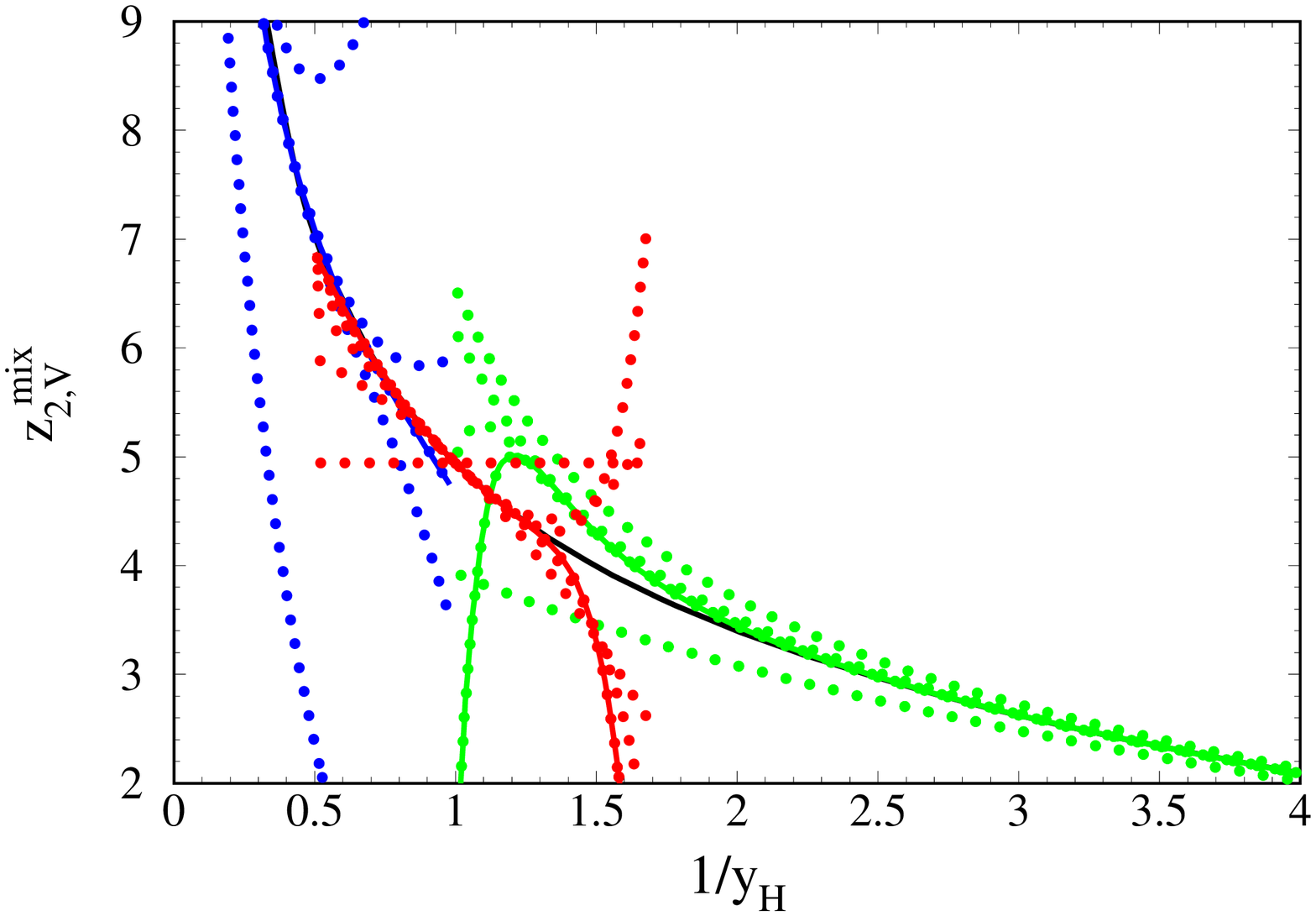,width=18em}
      }
      \\ \multicolumn{2}{c}{(c)}
    \end{tabular}
    \parbox{14.cm}{
      \caption[]{\label{fig::z2v}\sloppy
        (a) One-loop, (b) $1/\epsilon$ pole and (c) constant part of 
        the two-loop corrections to $Z_2^{V, \rm OS}$ as a
        function of $1/y_H=M_H/m_t$. The solid (coloured)
        lines correspond to the
        highest available order for each case. The dotted curves show 
        lower-order results and nicely demonstrate the convergence.
        The exact result, which is plotted (in black)
        over the whole $1/y_H$
        range is only visible in a small gap around $1/y_H\approx2$.}}
  \end{center}
\end{figure}

In Fig.~\ref{fig::z2v} we discuss our analytic results in numerical form.
Fig.~\ref{fig::z2v}(a) shows the finite part of the
one-loop contribution where the exact result is represented by the
solid line and the expansions are plotted as dotted curves. Similarly
to $z_m$ (cf. Fig.~\ref{fig::zm}) it can be seen that the dotted lines
nicely converge to the exact curve after including successively 
higher order expansion terms. As one can see, after taking into
account the result from the regions $y_H\to0,1$ and $\infty$ there
remains only a quite small range for $y_H$ ($1/y_H\approx 1.5 \ldots 2.0$)
where the (black) solid curve is still visible and the simple
expansions fail to provide good approximations.
The situation is very similar for the divergent $1/\epsilon$
and finite two-loop contribution which are shown in 
Figs.~\ref{fig::z2v}(b) and~\ref{fig::z2v}(c), respectively.
Thus we can conclude that it is possible to avoid the use of the
quite complicated exact expressions for $Z_2^V$ but to adopt an adequate
simple expansion. For Higgs boson masses outside the range
$M_H\approx 250 \ldots 300$~GeV one can simply use the corresponding
formula from Eq.~(\ref{eq::z2res}) while for 250~GeV$\le M_H \le$
300~GeV a straightforward interpolation provides a sufficiently good
approximation.

It is interesting to mention that the result obtained in the 
gaugeless limit approximates the full result within approximately 
20\% accuracy for Higgs boson masses between 100~GeV and 800~GeV.
For $z_m$ the situation is similar once the 
tadpole contributions are discarded. In case the latter are
included the relative deviation between the full result and the one in
the gaugeless limit becomes smaller~\cite{Faisst:2004gn}.

%- }}}
%- {{{ Conclusions:

\section{\label{sec::concl}Conclusions}

\indent

The renormalization constants constitute building blocks for the evaluation
of quantum corrections to various processes. 
In this paper the on-shell top quark mass and wave
function renormalization constants have been considered in the SM 
up to order $\alpha\alpha_s$. The inclusion of electroweak
effects introduces a further scale into the problem, as compared to
the QCD or QED corrections, which makes the calculation
of the integrals significantly more complicated.

We expressed $Z_m^{\rm OS}$ and $Z_2^{V/A,\rm OS}$ as a linear
combination of a handful master integrals which are known
analytically, however, contain quite involved functions.
For the complicated two-scale master integrals 
we applied the powerful method of asymptotic expansion in three
different kinematical regions, which leads to power expansions
multiplied by simple logarithms.
We could reproduce the result for $Z_m^{\rm OS}$ available in the literature. 
The expression for $Z_2^{V/A,\rm OS}$ is new and constitutes a
building block, e.g., in the mixed electroweak/QCD corrections for
top quark pair production at threshold.
We checked that both for $Z_m^{\rm OS}$ and $Z_2^{V/A,\rm OS}$ the
simple expansions agree very well with the exact result.
In particular, it has been shown
that the expansion for large top quark mass leads to compact formulae
which approximate the exact results quite precisely --- almost up to the 
point $m_t=M$ with $M=M_W, M_Z$ of $M_H$.
As far as the Higgs boson mass dependence is concerned,
also the expansions around $m_t=M_H$ and $m_t\ll M_H$ have been
considered.

%- }}}

%- {{{ Acknowledgements:

\vspace*{1em}

\noindent
{\large\bf Acknowledgements}\\
We would like the thank M. Faisst for useful discussions and for
providing expressions which we could use for cross checks. We thank
C.~Anastasiou for discussions about {\tt AIR},
A. Onishchenko, V. Smirnov and
O. Veretin for carefully reading the paper and helpful suggestions
and J. Piclum for discussions on the tadpole contributions.
This work was supported by the ``Impuls- und Vernetzungsfonds'' of the
Helmholtz Association, contract number VH-NG-008 and the SFB/TR~9.

%- }}}

\begin{appendix}

%- {{{ Analytic results for $z_m^{\rm OS}$:

\section{\label{app::zm}Analytic results for $z_m^{H, \rm ew}$ and
  $z_m^{H, \rm mix}$}

The exact dependence on $y_H$ of the one-loop electroweak
contribution reads
\begin{eqnarray}
  z_m^{H, \rm ew} &=& \frac{\yw^2}{16} \left[ \frac{3}{2\epsilon} +
    \frac{B_p(0,1)}{2 m_t^2} - \frac{B_p(1,0)}{2 m_t^2} - 2 \left(
    1-\frac{1}{4\yh^2} \right) B_p(1,1)   \right]
  \label{eq::zmew}
  \,,
\end{eqnarray}
where the function $B_p(n_1,n_2)$ corresponds to the one-loop on-shell
integral with a internal top quark and Higgs boson line
\begin{eqnarray}
  B_p(n_1,n_2) &=& \int \frac{{\rm d}^d k}{i \pi^{d/2}} \frac{e^{\epsilon
      \gamma_E}}{(k^2-M_H^2)^{n_1} (k^2+2 kq)^{n_2}}
  \,.
\end{eqnarray}
The special cases needed in this paper are given by

{\scalefont{0.8}
\begin{eqnarray}
  B_p(0,1) &=& m^2 \left( \frac{\mu^2}{m^2} \right)^{\epsilon}
  \left( \frac{1}{\epsilon} + 1 + \epsilon \left( 1 +\frac{\pi^2}{12}
  \right ) +\epsilon^2 \left( 1+\frac{\pi^2}{12}-\frac{\zeta(3)}{3}
  \right ) \right ) 
  \,,\nonumber\\
  B_p(1,0) &=& B_p(0,1)\bigg|_{m\to M}
  \,,\nonumber\\
  B_p(1,1) &=& \frac{\left(\frac{\mu^2}{M^2} \right)^{\epsilon} - (1-2
    \yh^2) \left(\frac{\mu^2}{m^2} \right )^{\epsilon}}{2\yh^2\epsilon}
  \left[ 1+2\epsilon+ \left(4+\frac{\pi^2}{12} \right)\epsilon^2 + \left( 8+
  \frac{\pi^2}{6}- \frac{\zeta(3)}{3}  \right ) \epsilon^3 \right]
  \nonumber \\ 
  && +  \left( \frac{\yh^4}{(4\yh^2-1)} \right)^{\epsilon} 
  \left( \frac{\mu^2}{m^2}\right)^{\epsilon}
  \left\{ \left[ 1+2 \epsilon+ \left( 4 + \frac{\pi^2}{12}
  \right)\epsilon^2 
  \right] \Psi_1 + 2 \epsilon \left(1 + 2\epsilon \right)
  \Psi_2 + 2 \epsilon^2 \Psi_3 \right\}\, , \nonumber \\
  \label{eq::Bpres}
\end{eqnarray}
}

\noindent
where we have included the order $\epsilon$ terms which are needed for
the two-loop expressions, and $\Psi_i$ is a shorthand notation for ($y=m/M$):
\begin{eqnarray}
  \Psi_i &=& \frac{1}{y}  \sqrt{4 - \frac{1}{y^2}} \left ( {\rm Ls}_i
  ( \pi ) - \frac{1}{2} 
      {\rm Ls}_i \left ( t_1 \right  ) - \frac{1}{2} {\rm Ls}_i \left
      ( t_2 \right 
      )  \right ) \, ,
      \label{eq::psi1}
\end{eqnarray}
with 
\begin{eqnarray}
  t_1 &=&  2 \arccos \left ( \frac{1}{2 y}  \right ) \,, \nonumber \\ 
  t_2 &=&  2 \arccos \left ( 1- \frac{1}{2 y^2} \right ) \,.
  \label{eq::t12}
\end{eqnarray}
The function ${\rm Ls}_i(z)$ is defined through
\begin{eqnarray}
  {\rm Ls}_i(z) &=& - \int_0^z {\rm d}x 
  \ln^{i-1} \left[2 \sin\left(  \frac{x}{2}
  \right ) \right]  \, .  
  \label{eq::lsdef}
\end{eqnarray} 

At two-loop order it is convenient to 
write the finite quantity $z_m^{H, \rm mix}$ in the form
\begin{eqnarray}
  z_m^{H, \rm mix} &=& a_m^{H, \rm mix} + b_m^{H, \rm mix}
  \label{eq::zmmix}
  \,,
\end{eqnarray}
where $a_m^{H,\rm mix}$ corresponds to the generic two-loop result
and $b_m^{H,\rm mix}$ origins from counterterm contributions and 
products of one-loop results. We obtain

{\scalefont{0.8}
\begin{eqnarray}
  a_m^{H, \rm mix} &=&
  -\frac{\ywbar^2}{64}\Bigg[
  -\frac{(2 - \epsilon + 3 \epsilon^2 + 4 \epsilon^3) (-1 + 4 \yhbar^2)}
  {2 \epsilon \yhbar^2}
  \frac{H_2}{\overline{m}_t^2}
  + (-1 + 2 \epsilon) 
  \frac{H_4}{\overline{m}_t^2}
  \nonumber\\&&\mbox{}
  - \frac{(1 + \epsilon) (3 - 3 \epsilon + 8 \epsilon^2)}{\epsilon} 
  \frac{H_1}{\overline{m}_t^4} 
  - \frac{(2 + \epsilon + 4 \epsilon^2 + 8 \epsilon^3) 
    (-1 + 4 \yhbar^2)}{\epsilon \yhbar^2} 
  H_5
  \nonumber\\&&\mbox{}
  + \frac{(-2 + \epsilon - 3 \epsilon^2 - 4 \epsilon^3 )
    + \yhbar^2\left(4 + 10 \epsilon - 2 \epsilon^2 + 20
  \epsilon^3\right)}
  {2 \epsilon \yhbar^2}  
  \frac{H_3}{\overline{m}_t^2}
  \nonumber\\&&\mbox{}
  - \frac{(-4 - \epsilon - 7 \epsilon^2 - 12 \epsilon^3 )
    + \yhbar^2\left( 12 \epsilon - 8 \epsilon^2 + 12 \epsilon^3 \right)}
  {2 \epsilon} 
  \frac{Y_2}{\overline{m}_t^4} 
  \nonumber\\&&\mbox{}
  + \frac{(-4 - \epsilon - 7 \epsilon^2 - 12 \epsilon^3 )
    + \yhbar^2\left(8 + 14 \epsilon + 6 \epsilon^2 + 36 \epsilon^3\right)}
  {2 \epsilon \yhbar^2} 
  \frac{Y_3}{\overline{m}_t^2}
  \nonumber\\&&\mbox{}
  + \frac{2 (-1 + \epsilon)^2}{\epsilon} 
  \frac{Y_1}{\overline{m}_t^2}
  - \frac{-2 + 4 \epsilon - 2\epsilon^2 
    + \yhbar^2\left(4 - 16 \epsilon + 8 \epsilon^2\right)}
  {\epsilon \yhbar^2} 
  Y_4
  \Bigg]
  \,, \nonumber\\
  b_m^{H, \rm mix} &=& \frac{\yw^2}{4}\Bigg\{
%  \nonumber\\&&\mbox{}
  \frac{-18 - 3 \epsilon + 15\epsilon^2}{32 \epsilon^2} - \frac{3}{32} \logmbar{\mu t}
  + \Bigg[
    - \frac{3(-1 + 2 \yhbar^2)}{32} \epsilon \logmbar{\mu t}^2  
    + \frac{(3 + \epsilon)(-1+2\yhbar^2)}{16} \logmbar{\mu t} 
    \nonumber\\&&\mbox{}
    - \frac{-18 - 16 \epsilon - 16 \epsilon^2 - \epsilon^2 \pi^2 
      + \yhbar^2\left(24 + 8 \epsilon + 32 \epsilon^2 
      + 2 \epsilon^2 \pi^2\right)}{64 \epsilon}
    \Bigg] \frac{B_p(1,0)}{\overline{m}_t^2}
  \nonumber\\&&\mbox{}
  + 
  \Bigg[
    \frac{3(-1 + 3 \yhbar^2)}{32 \yhbar^2} \epsilon \logmbar{\mu t}^2  
    + \frac{3 + \epsilon 
      + \yhbar^2 \left(-9 + 6 \epsilon\right)}{16 \yhbar^2} \logmbar{\mu t}
    \nonumber\\&&\mbox{}
    + \frac{-18 - 16 \epsilon - 16 \epsilon^2 - \epsilon^2 \pi^2 
      + \yhbar^2\left(108 + 24 \epsilon 
      + 3 \epsilon^2 \pi^2\right)}{64 \epsilon \yhbar^2}
    \Bigg] B_p(1,1)
  \nonumber\\&&\mbox{}
  + (-1 + 4 \yhbar^2)
  \Bigg[
    - \frac{3}{64 \yhbar^2} \epsilon \logmbar{\mu t}^2 
    + \frac{(3 + 4 \epsilon) \logmbar{\mu t} }{32 \yhbar^2} 
    - \frac{(16 + 32 \epsilon + \epsilon \pi^2)}{128 \yhbar^2}
    \Bigg]  \frac{y_H{\rm d}}{{\rm d} y_H} B_p(1,1)
  \Bigg\}
  \,.
\end{eqnarray}
}

\noindent
with $\ywbar$, $\yhbar$ and $\logmbar{\mu t}$ are
defined in Eq.~(\ref{eq::zmdef}).
The poles which are still present in $a_m^{H, \rm mix}$ and $b_m^{H, \rm mix}$
cancel in the proper sum.

Finally, the one- and two-loop coefficients determining 
the $\mu$ dependence are given by
\begin{eqnarray}
  z_m^{(1),H,\rm ew} &=& \frac{3\yw^2}{32}
  \,,\nonumber\\
  z_m^{(1),H,\rm mix} &=& \yw^2 \Bigg [ -\frac{3
      \sqrt{1-4\yhbar^2} \ln \left( 
      \overline{K}_H \right )}{256 \yhbar^4} (8\yhbar^2+1) - \frac{3
      \logmbar{H}}{256 \yhbar^4} 
    (6\yhbar^2+1) 
  \nonumber\\&&\mbox{}
    + \frac{3}{128\yhbar^2} (21\yhbar^2+1)  \Bigg ]
  \,,\nonumber\\
  z_m^{(2),H,\rm mix} &=&  -\frac{9\yw^2}{64}
  \, ,
\end{eqnarray}
where $\overline{K}_H = \frac{1-\sqrt{1-4\yhbar^2}}{1+\sqrt{1-4\yhbar^2}}$.

%- }}}
%- {{{ Analytic results for $Z_2^{\rm OS}$:

\section{\label{app::z2}Analytic results for $Z_{2,V}^{H, \rm ew}$ and
  $Z_{2,V}^{H, \rm mix}$}

The one-loop contribution from the Higgs boson exchange diagram
to $Z_{2,V}^{H,\rm ew}$ reads
\begin{eqnarray}
  Z_{2,V}^{H,\rm ew} &=& -\frac{\yw^2}{32} \left[ 4 (1-\epsilon) \yh^2
    \frac{B_p(1,0)}{m_t^2}
    + \frac{3-2 \epsilon}{\yh^2}  B_p(1,1)
    \right.\nonumber\\&&\left.\mbox{}
    + \frac{3-2\epsilon}{m_t^2} \left( B_p(0,1)-B_p(1,0)-2 m_t^2 B_p(1,1)
    \right)  \right] \,,
  \label{eq::z21lhexact}
\end{eqnarray} 
which --- after inserting the results for $B_p(n_1,n_2)$ from 
Eq.~(\ref{eq::Bpres}), with $m=m_t$ and $M=M_H$,
and expanding in $\epsilon$ --- can be cast in the form
\begin{eqnarray}
  f_{2,V}^{H,\rm ew} &=&  - \frac{\yw^2}{32} \,,\nonumber\\
  z_{2,V}^{H,\rm ew} &=&  - \yw^2 \Bigg [ \frac{3 \sqrt{1-4\yh^2} \ln \left(
      K_H \right)}{64\yh^4} ( 2 \yh^2 - 1 ) +  \frac{\logm{H}}{64 \yh^4}
    ( - 8 \yh^4 + 12\yh^2 - 3  ) \nonumber \\
    && +  \frac{( -7 \yh^2 + 3 )}{32\yh^2} \Bigg ]
  \,,
  \label{eq::f2z2}
\end{eqnarray}
where $y_H$ and $\logm{H}$ are defined in Eq.~(\ref{eq::z2def})
and $K_H=\frac{1-\sqrt{1-4\yh^2}}{1+\sqrt{1-4\yh^2}}$.

At two-loop order we again split $Z_{2,V}^{H,\rm OS}$ into two parts,
\begin{eqnarray}
  Z_{2,V}^{H,\rm mix} &=& A_2^{H, \rm mix} + B_2^{H, \rm mix}
  \,,
\end{eqnarray}
where $A_2^{H, \rm mix}$ corresponds to the generic two-loop result
and $B_2^{H, \rm mix}$ origins from counter\-term contributions and
squared one-loop results. We obtain

{\scalefont{.75}
\begin{eqnarray}
  A_2^{H, \rm mix} &=& - \frac{\yw^2}{64} \Bigg\{
    -\frac{(9  -  \epsilon  + 12 \epsilon^2  + 26 \epsilon^3)- ( 30
      + 28 \epsilon  + 22 \epsilon^2   + 100 \epsilon^3 ) \yh^2 + (56
      \epsilon + 16 \epsilon^2 - 72 \epsilon^3) \yh^4 }{(4 \yh^2
      - 1) \epsilon } \frac{H_1}{m_t^4} \nonumber \\
    &&+ \frac{ (- 6  + 5 \epsilon - 13 \epsilon^2 + 2 \epsilon^3) + (14  -
      22 \epsilon  + 96 \epsilon^2  - 160 \epsilon^3)\yh^2}{2 \epsilon
      \yh^2} \frac{H_2}{m_t^2} \nonumber \\ 
    && 
    + \Bigg[
      \frac{ 
        (  - 6 + 5 \epsilon - 13 \epsilon^2 + 2 \epsilon^3 )
        + ( 30  + 58 \epsilon^2 - 16 \epsilon^3 )\yh^2 }
        { 2 (4 \yh^2 - 1) \epsilon \yh^2} 
        \nonumber\\&&\mbox{}
      + \frac{( - 16 - 44 \epsilon - 12 \epsilon^2 + 56 \epsilon^3) \yh^2 +
        (32 \epsilon - 160 \epsilon^3 ) \yh^4}
      { (4 \yh^2 - 1) \epsilon } 
      \Bigg]
      \frac{H_3}{m_t^2} 
    \nonumber \\ 
    &&  - \frac{-2 + 20 \epsilon^2 - 54 \epsilon^3 }{\epsilon}
    \frac{H_4}{m_t^2} + \frac{ - ( 6 + 12 \epsilon^2 + 14 \epsilon^3) + (14 - 12
      \epsilon  +  72 \epsilon^2  - 72 \epsilon^3) \yh^2}{\epsilon
  \yh^2 } 
    H_5 \nonumber \\
    && +\frac{ (6 - 16 \epsilon  + 14 \epsilon^2  - 4 \epsilon^3)+(-20 + 58 \epsilon - 56 \epsilon^2 + 16 \epsilon^3) \yh^2}{ (4 \yh^2 -
      1)\epsilon } \frac{Y_1}{m_t^2} \nonumber \\
    && -\frac{ ( - 12 - 5 \epsilon - 13 \epsilon^2 - 46 \epsilon^3)+(40  +
  88 \epsilon  - 68 \epsilon^2  + 380 \epsilon^3)\yh^2 + (-200
  \epsilon + 352 \epsilon^2 - 744 \epsilon^3) \yh^4 }{2(4\yh^2 -
  1)\epsilon}  \frac{Y_2}{m_t^4} 
    \nonumber \\  && 
    +\Bigg[
     \frac{- ( 12 + 3 \epsilon + 11 \epsilon^2 + 54 \epsilon^3) + (60  + 54
      \epsilon - 6 \epsilon^2  + 348 \epsilon^3 )\yh^2}
      {2 (4 \yh^2 - 1)\epsilon \yh^2}
      \nonumber\\&&\mbox{}
      +\frac{ ( - 32 - 64\epsilon + 68 \epsilon^2  - 236 \epsilon^3) \yh^2 
      + (24 \epsilon - 40\epsilon^2 + 16 \epsilon^3 )\yh^4 }
    {(4 \yh^2 - 1)\epsilon}
    \Bigg]\frac{Y_3}{m_t^2} 
    \nonumber \\ && 
    -\frac{ (-6 + 16 \epsilon - 14 \epsilon^2 + 4 \epsilon^3) + ( 30 -
  96 \epsilon + 88 \epsilon^2 - 24 \epsilon^3) \yh^2 + (-32 + 144
  \epsilon -  128 \epsilon^2  + 32 \epsilon^3) \yh^4}{(4 \yh^2 - 1)
  \epsilon \yh^2} Y_4 \, \Bigg\} \, ,
  \nonumber\\
  \label{eq::z2ew}
\end{eqnarray}
\begin{eqnarray}
  B_2^{H, \rm mix} &=& \frac{\yw^2}{4} \Bigg \{ \Bigg [
    \frac{ 21+10\epsilon - (108+36\epsilon) \yh^2 +
      (24+8\epsilon) \yh^4}{16 (4 \yh^2 - 1)} \logm{\mu t}  -
    \frac{3 \epsilon (7-36\yh^2+8\yh^4 )}{32 (4 \yh^2 - 1)} \logm{\mu
      t}^2 \nonumber \\ 
    && + \frac{-(21 + 10 \epsilon + 38 \epsilon^2) + ( 108 + 36 \epsilon +
      180 \epsilon^2) \yh^2 - ( 24 + 8 \epsilon + 32 \epsilon^2)\yh^4}{16
      (4\yh^2 - 1) \epsilon} + \frac{( -7 + 36 \yh^2 - 8  \yh^4
    )\epsilon \pi^2 }{64 (4\yh^2 - 1)} \Bigg ]
  \nonumber \\ && \mbox{}
  \times \frac{B_p(0, 1)}{m_t^2}
  + \Bigg [\frac{ - ( 21 + 10 \epsilon) + ( 132 + 44 \epsilon)
      \yh^2 - (156+28 \epsilon) \yh^4}{16 (4 \yh^2 - 1)} \logm{\mu t}
    - \frac{3 \epsilon \, (-7 + 44\yh^2 - 52 \yh^4)}{32 (4 \yh^2 - 1)}
    \logm{\mu t}^2  \nonumber \\ 
    && + \frac{21 + 10\epsilon + 38 \epsilon^2 - (  132  + 44
      \epsilon  + 212 \epsilon^2 )\yh^2 + (156 + 28 \epsilon  + 200
      \epsilon^2)\yh^4 }{16 \epsilon (4 \yh^2 - 1)} 
  \nonumber \\ && 
    - \frac{(-7 + 44 \yh^2 - 52
      \yh^4 ) \epsilon \pi^2 }{64 (4 \yh^2 - 1) } \Bigg ]
  \frac{B_p(1, 0)}{m_t^2}  
  + \Bigg [\frac{3 \epsilon \, ( -7 + 52 \yh^2 - 92  \yh^4 + 8  \yh^6
      )}{32 (4 \yh^2 - 1) \yh^2 }  \logm{\mu t}^2  
    \nonumber\\&&\mbox{}
    +  \frac{ 21  + 10
      \epsilon - ( 156  + 52 \epsilon)\yh^2  + ( 276 + 68 \epsilon)\yh^4  +
      ( - 24  + 16 \epsilon)\yh^6 }{16 (4 \yh^2 -  1) \yh^2}  \logm{\mu
      t} \nonumber \\ 
    && +\frac{ -( 21 + 10 \epsilon + 38 \epsilon^2 )+ ( 156  + 52 \epsilon
      + 268 \epsilon^2 ) \yh^2 - (276 + 68 \epsilon + 456 \epsilon^2) \yh^4 +
      (24 - 16 \epsilon) \yh^6 }{16 (4 \yh^2 - 1)\epsilon \yh^2  }
    \nonumber \\ && + \frac{
      ( -7 + 52 \yh^2 - 92 \yh^4  + 8 \yh^6 ) \epsilon \pi^2 }{64 (4 \yh^2 -
      1) \yh^2 } \Bigg ] B_p(1, 1) \Bigg \}
  \,.
  \label{eq::z2mix}
\end{eqnarray}
}

Note that the individual contributions  $A_2^{H, \rm mix}$ and
$B_2^{H, \rm mix}$ still depend on the QCD
gauge parameter $\xi$ which 
cancels in the proper sum. For simplicity these terms have already been omitted in 
Eqs.~(\ref{eq::z2ew}) and~(\ref{eq::z2mix}).

From the formulae~(\ref{eq::z2ew}) and~(\ref{eq::z2mix}) 
it is straightforward to extract exact expressions
for the pole parts which are given by
\begin{eqnarray}
  g_{2,V}^{\rm mix} &=& \frac{3 \yw^2}{64}
  \,,\nonumber\\
  h_{2,V}^{H,\rm mix} &=& \yw^2 \Bigg [ -\frac{\ln \left(K_H \right)}{\yh^4}
    \sqrt{1-4\yh^2} \left( \frac{7}{256} 
    + \frac{1}{128(1-4\yh^2)}  \right )  - \frac{9 \logm{H}}{256\yh^4}
    \nonumber \\  
    &&  +\frac{\ln \left( K_H \right )}{\yh^2} \sqrt{1-4\yh^2}
    \left(\frac{11}{128} +\frac{1}{64(1-4\yh^2)}  \right )
    + \frac{9\logm{H} }{64 \yh^2} + \frac{9}{128 \yh^2}  \nonumber \\
    && -\frac{3}{32} \logm{H} + \frac{\ln
      \left (K_H \right )}{16 \sqrt{1-4\yh^2}}  -\frac{7}{128} \Bigg
  ]\,,
  \label{eq::z2Hpoles}
\end{eqnarray}
where $K_H$ is defined after Eq.~(\ref{eq::f2z2}).

For completeness we want to provide the exact result covering the
renormalization scale dependence for $Z_2^{V,H,\rm OS}$. The corresponding
coefficients are defined in analogy to the ones for $z_m$ in
Eq.~(\ref{eq::zmdef}) and read
\begin{eqnarray}
  z_{2,V}^{(1),H,\rm ew} &=& \frac{\yw^2} {32}
  \,, \nonumber\\
  h_{2,V}^{(1),H,\rm mix} &=& -\frac{3\yw^2}{32}
  \,, \nonumber\\
  z_{2,V}^{(1),H,\rm mix} &=& \yw^2 \Bigg [ \frac{\sqrt{1-4\yh^2}\ln
      \left(K_H \right )}{128 \yh^4} (-22\yh^2+7) - \frac{\ln
      \left( K_H \right ) }{ 64 \yh^4 \sqrt{1-4\yh^2} } (8\yh^4+2\yh^2-1)
    \nonumber \\
    && + \frac{3 L_H}{128\yh^4} (  8 \yh^4 - 12\yh^2 + 3 ) +
    \frac{(7\yh^2-9 )}{64\yh^2} \Bigg ] 
  \,, \nonumber\\
  z_{2,V}^{(2),\rm mix} &=& \frac{3\yw^2}{32}
  \,.
  \label{eq::z2Hmu}
\end{eqnarray}

%- }}}
%- {{{ Master integrals:

\section{\label{app::ae_of_masters}Master integrals}

\indent

In this Section we discuss the master integrals occuring in our calculation.
The master integrals needed for the neutral boson exchange
which reduce to products of one-loop integrals,
to two-loop one-scale integrals or to two-loop vacuum integrals
are given by (see, e.g., Refs.~\cite{Fleischer:1999tu,Davydychev:2000na})

{\scalefont{1.0}
\begin{eqnarray}
  H_1 &=& m^4 \left(\frac{\mu^2}{m^2}\right)^{2 \epsilon} 
  \left[ \frac{1}{{\epsilon}^2} +
  \frac{2}{{\epsilon}} +3+\frac{\pi^2}{6} + {\epsilon} \left (
  4-\frac{2}{3}  \zeta(3)+\frac{\pi^2}{3} \right ) 
  + {\cal O}\left(\epsilon^2\right) \right]
  \,,  \nonumber   \\
  H_2 &=&  M^2 \left(\frac{\mu^2}{M^2} \right )^{2 \epsilon}
  \left[ \frac{1}{2 {\epsilon}^2} +
  \frac{3}{2\epsilon} +\frac{7}{2}+\frac{\pi^2}{4} + {\epsilon} \left
  ( \frac{15}{2}-\frac{4}{3} \zeta(3)+\frac{3\pi^2}{4} \right ) 
  + {\cal O}\left(\epsilon^2\right) \right]
  \,, \nonumber  \\
  H_3 &=& m^2  \left ( \frac{\mu^2}{m^2} \right )^{2 \epsilon}
  \left\{ \frac{1}{{\epsilon}^2} \left ( 1 + \frac{1}{2 y^2}\right ) +
  \frac{1}{\epsilon} \left ( 3 + \frac{3}{2 y^2} + \frac{1}{y^2}
  \ln \left( y^2 \right) \right ) 
  \right. \nonumber   \\
  && \left.  + \left ( 7 + \frac{\pi^2}{6} \right ) \, \left ( 1 +
  \frac{1}{2 y^2} \right ) + \frac{3}{y^2} \ln \left (
  y^2 \right )
   + \frac{1}{2 y^2} \ln^2 \left( y^2 \right ) +  2 \Omega_2
  \right. \nonumber \\ 
 && \left. + \epsilon \left[ \left ( 15 + \frac{\pi^2}{2} - \frac{2
  \zeta(3)}{3} \right ) \, \left ( 1+ \frac{1}{2 y^2} \right ) + \left
  ( 7 + \frac{\pi^2}{6} \right ) \frac{1}{y^2} \ln \left (
  y^2 \right ) + \frac{3}{2 y^2} \ln^2 \left (
  y^2 \right )  \right. \right. \nonumber \\
 &&  \left. \left. + \frac{1}{6 y^2} \ln^3 \left (
  y^2 \right ) + \left (6-2
  \ln \left (  \frac{4}{y^2} - \frac{1}{y^4}  \right ) \right )
  \Omega_2  + 2\Omega_3  \right] + {\cal O}\left(\epsilon^2 \right)
  \right\} 
  \,,  \nonumber
\end{eqnarray}
\begin{eqnarray}
  Y_1 &=&  
  m^2 \left ( \frac{\mu^2}{m^2} \right )^{2 \epsilon} 
  \left[ \frac{3}{2 {\epsilon}^2} +
  \frac{17}{{4 \epsilon}}+\frac{59}{8}+\frac{\pi^2}{4} + {\epsilon}
  \left ( \frac{65}{16}-  \zeta(3)+\frac{49 \pi^2}{24} \right ) 
  + {\cal O}\left(\epsilon^2\right) \right]
  \,,  \nonumber  \\
  Y_2 &=& y^2 M^4 \left ( \frac{\mu^2}{m^2} \right
  )^{-\epsilon} \left ( \frac{\mu^2}{M^2} \right )^{\epsilon} 
  \frac{H_1}{m^4}
  \,,  \nonumber   \\
  Y_3 &=& B_p(0,1) B_p(1,1)\,,
\end{eqnarray}
}

\noindent
with $y=m/M$ and
\begin{eqnarray}
  \Omega_i &=& \Psi_i 
  - \frac{1}{2y} \sqrt{4-\frac{1}{y^2}} {\rm Ls}_i \left(t_1\right)\,,
\end{eqnarray}
where $\Psi_i$, $t_1$ and $t_2$ are defined in Eqs.~(\ref{eq::psi1}) 
and~(\ref{eq::t12}).

As already mentioned in the main text, analytical results for the 
complicated two-scale master integrals can be found in
Ref.~\cite{Jegerlehner:2003py}. We refrain from repeating them here.
Instead we perform expansions in the 
limits $y\to0$, $y\to1$ and $y\to\infty$ which we derived
independently with the
help of asymptotic expansions~\cite{Smirnov:pj}.
In many phenomenological applications it is advantageous to use 
the handy approximation formulae in favour of
the complicated exact expressions.

The expansions in the three different limits require a
strategy of its own. In the case of a large boson mass $M$ one obtains,
next to on-shell, also vacuum integrals up to two loops (both for charged and
neutral boson exchange) which are, e.g.,
implemented in MATAD~\cite{Steinhauser:2000ry}. 
The subdiagrams contributing in
this limit can be obtained in a completely automated way
with the help of {\tt exp}~\cite{Seidensticker:1999bb,Harlander:1997zb}. 

In the limit $y\to1$, the case of a neutral boson exchange reduces to
a simple Taylor expansion. The resulting 
integrals are well studied within QCD and documented in a 
program code~\cite{Fleischer:1999tu}.
The expansion for $y\to1$ of the diagrams involving a $W$ boson is
more complicated, which is due to the appearance of additional massless
particles, the bottom quarks. We did not perform the calculation of the
diagrams in this limit since for the practical applications only the
limit $m_t\gg M_W$ is needed.

Finally for the case $y\to\infty$, which is the limit 
of main interest, a careful inspection of the
regions~\cite{Beneke:1997zp,Smirnov:1998vk} contributing to the 
integral under consideration is in order. 

In the following we present a pedagogical example which illustrates
the procedure in more detail.
Let us consider the master integral $Y_4$ defined through
\begin{eqnarray}
  Y_4 &=& \frac{e^{2\epsilon\gamma_E}}{(i\pi^{d/2})^2} 
  \int\frac{ {\rm d}^d k {\rm d}^d l }{(k^2+2kq) \, (l^2-2lq) \,
  ((k-l)^2+2q(k-l)) \,  (l^2-M^2)} \,.
\end{eqnarray}
In the limit $y\to0$, i.e. $m\ll M$ one has to consider the
cases 
(i) $|\vec{k}|\sim|\vec{l}|\sim M$, 
(ii) $|\vec{k}|\sim m$, $|\vec{l}|\sim M$, 
(iii) $|\vec{k}|\sim|\vec{l}|\sim m$, and 
(iv) $|\vec{k}|\sim|\vec{l}|\sim M$ but with 
$|\vec{k} - \vec{l}|\sim m$, 
which we denote as hard-hard (HH), soft-hard (SH), soft-soft (SS) and
hard-soft (HS) region. 
In each region one has to perform a simple Taylor
expansion of each propagator in the small quantities.
E.g., if $\vec{k}$ is hard the propagator $1/(k^2+2kq)$ is expanded in
$2kq/k^2$ since we have $q^2=m^2$.
After adding the contributions from all regions one finally obtains the
result for $Y_4$
\begin{eqnarray}
  Y_{4,0} &=& Y_4({\rm HH}) + Y_4({\rm SH}) + Y_4({\rm SS}) + Y_4({\rm HS})
  \,,
\end{eqnarray}
where the integrals in the individual region have the form
\begin{eqnarray}
  Y_4({\rm HH})  &=& \sum_{n_1,...,n_5} {\mathcal
    C}_{n_1...n_5}^{{\rm HH}} \int\frac{ {\rm d}^d k {\rm d}^d l \, \,
    (-2kq)^{n_1} \, (-2lq)^{n_2}}{(k^2)^{n_3} \, ((k-l)^2-M^2)^{n_4} \,
    (l^2)^{n_5}} \,,  \nonumber \\
  Y_4({\rm SH}/{\rm HS})  &=& \sum_{n_1,...,n_5} {\mathcal
    C}_{n_1...n_5}^{{\rm SH/HS}}\int\frac{ {\rm d}^d k {\rm d}^d l \, \,
    (-2kq)^{n_1} \, (-2kl)^{n_2} \, 
    (-2lq)^{n_3}}{(k^2-M^2)^{n_4}  \, (l^2+2lq)^{n_5}} \,,  \nonumber \\
  Y_4({\rm SS})  &=& \sum_{n_1,...,n_5} {\mathcal
    C}_{n_1...n_5}^{{\rm SS}} \int\frac{ {\rm d}^d k {\rm d}^d l \, \,
    (k^2)^{n_1} \, (l^2)^{n_2}}{(k^2+2kq)^{n_3}  \,
    (l^2-2lq)^{n_4} ((k-l)^2+2q(k-l))^{n_5} } \, .  
\end{eqnarray}
We have used partial fractioning and rearranged the different
propagators in a convenient way. The indexes $n_1$ to $n_5$ run
over appropriate 
integer values, and the ${\mathcal C}^{ab}_{n_1...n_5}$ are shorthand 
notations for a set of functions depending on
both scales $M$ and $m$ and the dimension $d$.
The first four terms of $Y_4$ expanded for  $y\to0$ read
\begin{eqnarray}
  Y_{4,0} &=& \left( \frac{\mu^2}{m^2} \right )^{2\epsilon} \left\{
  \frac{1}{2\epsilon^2} 
  + \left[\frac{3}{2} + \ln(y^2) + \left(
  \frac{1}{2}+ \ln(y^2) \right ) y^2 + \left( \frac{5}{3}+ 2\ln(y^2)
  \right) y^4 
  \right.\right. \nonumber \\&& \left. \left.
  + \left( \frac{59}{12} + 5\ln(y^2) \right) y^6 \right]
  \frac{1}{\epsilon} 
  + \frac{7}{2}+ \frac{\pi^2}{4} + 3\ln(y^2) + \ln^2(y^2) +
  \right. \nonumber \\ && \left.
  \left( 2 -\frac{\pi^2}{6} + \frac{11\ln(y^2)}{2}  \right ) y^2
  + \left ( \frac{85}{9} - \frac{\pi^2}{3} +
  \frac{26\ln(y^2)}{3}  \right ) y^4 
  \right. \nonumber \\
  && \left. 
  + \left ( \frac{244}{9} - \frac{5\pi^2}{6} + \frac{241\log(y^2)}{12}
  \right ) y^6 \right\} + {\cal O}(y^8)
  \,,
  \label{eq::Y40}
\end{eqnarray}
where we refrain from listing the contributions of order $\epsilon$.

In the region $y\to\infty$, where $M\ll m$ a
straightforward inspection reveals only two different ranges for the
integration momenta ${\vec k}$ and ${\vec l}$. In
the first one (hh) both momenta are of order $m$, which now
is hard; for the second one (hs) ${\vec k}$ is hard
and ${\vec l}$ is soft, i.e. of order $M$. Thus $Y_4$ can be written as
\begin{eqnarray}
  Y_4 &=&  Y_4({\rm hh}) + Y_4({\rm hs}) \,, 
\end{eqnarray}
where the following integrals occur
\begin{eqnarray}
  Y_4({\rm hh})  &=& \sum_{n_1,...,n_5}{\mathbb
    C }_{n_1...n_5}^{{\rm hh}} \int\frac{ {\rm d}^d k {\rm
      d}^d l \, (M^2)^{n_5} }{(k^2+2kq)^{n_1} \, (l^2-2lq)^{n_2} \,
    ((k-l)^2+2q(k-l))^{n_3} \, (l^2)^{n_4} } \,,  \nonumber \\ 
  Y_4({\rm hs})  &=& \sum_{n_1,...,n_5} {\mathbb
    C}_{n_1...n_5}^{{\rm hs}} \int\frac{ {\rm d}^d k {\rm d}^d l \, \,
    (2kl)^{n_1} \, (-2lq)^{n_2} }{(k^2+2kq)^{n_3} \, (k^2)^{n_4} \,
    (l^2-M^2)^{n_5}} \, .  \nonumber \\ 
\end{eqnarray} 
Note that the collection of integrals in $Y_4({\rm hh})$ and the
ones in $Y_4({\rm SS})$ are equivalent, and, at the
same time, equal to the ones one would need to solve in order to get an
arbitrary two loop ${\mathcal{O}}(\alpha_s^2)$ strong QCD correction. 
The first seven terms of $Y_4$ expanded for $y\to\infty$ read
\begin{eqnarray}
  Y_{4,\infty} &=& \left( \frac{\mu^2}{m^2} \right )^{2\epsilon} \left\{
  \frac{1}{2\epsilon^2} 
  + \left[ \frac{5}{2}-\frac{\pi}{y} +
  \left(1+\frac{\ln(y^2)}{2} \right) \frac{1}{y^2}  + \frac{\pi}{8 y^3}
  - \frac{1}{12 y^4} + \frac{\pi}{128 y^5} 
  \right.\right. \nonumber \\  && \left.\left. 
  - \frac{1}{120 y^6}  \right] \frac{1}{\epsilon} 
  + \frac{19}{2} - \frac{7\pi^2}{12} + \left (-2 + 2 
  \logtwo - \ln(y^2) \right )
  \frac{\pi}{y} 
  \right. \nonumber \\&& \left. 
  + \left( -3 + \frac{\pi^2}{6} + \frac{\ln^2(y^2)}{4}
  \right ) \frac{1}{y^2} 
  - \left( \frac{1}{6} + \frac{\logtwo}{4} -
  \frac{\ln(y^2)}{8} \right ) \frac{\pi}{y^3} + \left( \frac{2}{3}+
  \frac{\ln(y^2)}{6}  \right ) \frac{1}{y^4} 
  \right. \nonumber \\  && \left.
  + \left( \frac{19}{960} -
  \frac{\logtwo}{64} + \frac{\ln(y^2)}{128} \right ) \frac{\pi}{y^5}
  + \left( \frac{131}{3600} + \frac{\ln(y^2)}{60}   \right )
  \frac{1}{y^6}  + {\cal O}\left(\frac{1}{y^7}\right)\right\}
  \,.
  \label{eq::Y4infty}
\end{eqnarray}

Finally, let us consider limit the $y\to 1$. Here the asymptotic
expansion leads to only one region which corresponds to a naive Taylor
expansion in the quantity $\Delta= (m^2-M^2)$. This leads to 
the integrals of the type
\begin{eqnarray}
  Y_4 &=&  \sum_{n_1}
  {\bf
    C}_{n_1} 
  \int\frac{ {\rm d}^d k {\rm d}^d l \, (\Delta)^{n_1-1} }{(k^2+2kq)
    \, (l^2-2lq) \, ((k-l)^2+2q(k-l))  \,  (l^2-m^2)^{n_1} } \,.
\end{eqnarray}
The first six terms in the expansion are given by

{\scalefont{0.8}
\begin{eqnarray}
  Y_{4,1} &=& \left( \frac{\mu^2}{m^2} \right )^{2\epsilon} \left\{
  \frac{1}{2\epsilon^2} +  \left[ \frac{5}{2}-\frac{\pi\sqrtthree}{3}
  + \frac{\pi\sqrtthree}{9} \yone + \left( -\frac{1}{6} +
  \frac{2\pi\sqrtthree}{27} \right ) \yone^2 +
  \frac{2\pi\sqrtthree}{81}\yone^3  + \frac{4\pi\sqrtthree}{243}
  \yone^4 
  \right. \right. \nonumber \\  && \left. \left. \mbox{}
  + \left( \frac{1}{180} + \frac{8\pi\sqrt3}{729} \right ) \yone^5
  \right] \frac{1}{\epsilon} 
  +\frac{19}{2}
  + \left(\frac{\logthree}{3}-\frac{4}{3} \right) \pi\sqrtthree
  -\frac{63}{4} S_2 
  + \left[
  -\frac{\pi^2}{36}+ \frac{21}{4} S_2 
  \right.\right. \nonumber \\&& \left.\left. \mbox{} 
  - \left(  \frac{\logthree}{9} + \frac{1}{9} \right ) \pi
  \sqrtthree \right] \yone 
  + \left[ -\frac{1}{3} + \frac{7}{2} S_2 + \left (
  -\frac{2\logthree}{27} + \frac{1}{54} \right ) \pi
  \sqrtthree \right] \yone^2 + \left[ -\frac{11}{54} + \frac{7}{6} S_2
  \right.\right. \nonumber \\&& \left.\left. \mbox{} 
  + \left(-\frac{2\logthree}{81} + \frac{23}{486} \right ) \pi
  \sqrtthree \right] \yone^3  
  + \left[ -\frac{25}{162} + \frac{7}{9} S_2 + \left(
  -\frac{4\logthree}{243} + \frac{119}{2916} \right )
  \pi \sqrtthree \right] \yone^4 \right. \nonumber \\
  && \left.\mbox{} + \left[ -\frac{473}{4860} +
  \frac{14}{27} S_2 + \left( -\frac{8\logthree}{729} +
  \frac{1363}{43740}  \right ) \pi \sqrtthree \right] \yone^5 
  + {\cal O}(\yone^6)
  \right\}
  \,,
  \label{eq::Y41}
\end{eqnarray}
}

\noindent
with $\yone=1-1/y^2=\Delta/m^2$ and $S_2$ as defined in 
Eq.~(\ref{eq::const_def}).

We proceeded in an analog way for the remaining two integrals in
Eq.~(\ref{eq::master_neutral}). Furthermore, for the seven integrals in
Eqs.~(\ref{eq::master_charged1}) and~(\ref{eq::master_charged})
we evaluated the phenomenological limit
$y\to \infty$. However, we refrain from listing the results
explicitly. 

%- }}}

\end{appendix}

%- {{{ bibliography:

%- }}}

\end{document}